\documentclass[prb,twocolumn,amsmath,amssymb]{revtex4}

\usepackage{graphicx}
\usepackage{dcolumn}
\usepackage{bm}

\begin{document}


\title{Self-consistent perturbation expansion for Bose-Einstein condensates\\
satisfying Goldstone's theorem and conservation laws}

\author{Takafumi Kita}
\affiliation{Department of Physics, Hokkaido University, Sapporo 060-0810, Japan}%

\date{\today}

\begin{abstract}
Quantum-field-theoretic descriptions of interacting condensed bosons have suffered from the lack of self-consistent approximation schemes satisfying Goldstone's theorem and dynamical conservation laws simultaneously. We present a procedure to construct such approximations systematically by using either an exact relation for the interaction energy or the Hugenholtz-Pines relation to express the thermodynamic potential in a Luttinger-Ward  form. Inspection of the self-consistent perturbation expansion up to the third order with respect to the interaction shows that the two relations yield a unique identical result at each order, reproducing the conserving-gapless mean-field theory [T.\ Kita, J.\ Phys.\ Soc.\ Jpn.\ {\bf 74},\ 1891 (2005)] as the lowest-order approximation. The uniqueness implies that the series becomes exact when infinite terms are retained. We also derive useful expressions for the entropy and superfluid density in terms of Green's function and a set of real-time dynamical equations to describe thermalization of the condensate.
\end{abstract}

\maketitle

\section{Introduction}

Broken symmetry and self-consistency are among the most fundamental concepts 
in modern theoretical physics.
The former brings a drastic change in the system with the appearances of 
the order parameter  \cite{Anderson84,PS95} and the corresponding Nambu-Goldstone boson. \cite{PS95,Nambu61,Goldstone61}
The order parameter has to be determined self-consistently 
together with quasiparticles responsible for the excitation. 
Thus, the latter concept is also essential for describing broken symmetry phases.

Self-consistency plays crucial roles even in normal systems
as exemplified in the Landau theory of Fermi liquids \cite{Landau56,BP91} where
an external perturbation produces a self-consistent molecular field 
to yield an enhanced response in some cases. 
It is also adopted commonly in various practical approximation schemes such as the Hartree-Fock theory
and the density-functional theory;  \cite{Kaxiras03} with some infinite series incorporated in terms of the latter,
self-consistent approximations can be far more effective than the simple perturbation expansion.
It is worth pointing out that the Landau theory of Fermi liquids has been 
justified microscopically with the self-consistent quantum field theory,  \cite{AGD63}
which in turn enabled Leggett to extend the Landau theory to superfluid Fermi liquids. \cite{Leggett65}

Among those self-consistent approximations is Baym's $\Phi$-derivable approximation
with a unique property 
of obeying various dynamical conservation laws automatically. \cite{Baym62}
This is certainly a character indispensable for describing
nonequilibrium phenomena but not met by the simple perturbation expansion.
Indeed, the $\Phi$-derivable approximation seems the only systematic microscopic scheme within the quantum field theory
which enables us to study equilibrium and nonequilibrium phenomena on an equal footing.
It includes the Hartree-Fock theory and the Bardeen-Cooper-Schrieffer (BCS) theory of superconductivity
as notable examples. Moreover, the Boltzmann equation can be derived as a special case of the $\Phi$-derivable
approximation. \cite{Baym62,KB62}

The key functional $\Phi=\Phi[G]$ above was introduced by Luttinger and Ward \cite{LW60}  as part of the exact equilibrium thermodynamic potential for the normal state in terms of the Matsubara Green's function $G$. 
Based on a self-consistent perturbation expansion,
the expression also provides a systematic approximation scheme
with the desirable property of including the exact theory as a limit. \cite{Luttinger60} 
Indeed, Luttinger \cite{Luttinger60} subsequently used the Luttinger-Ward functional to obtain some general results on normal Fermi systems such as the Fermi-surface sum rule anticipated by Landau.  \cite{Landau56}
Note that nonequilibrium systems can be handled with essentially the same techniques as equilibrium cases by a mere change from the imaginary-time Matsubara contour into the real-time Keldysh contour.  \cite{Keldysh64,CSHY85,RS86}

Thus, it would be useful to have a practical $\Phi$-derivable approximation of Bose-Einstein condensates (BECs)
with broken U(1) symmetry,
where quite a few dynamical experiments have been carried out \cite{PS08,PS03,GNZ09} since the realization of the Bose-Einstein condensation with a trapped atomic gas in 1995. \cite{AEMWC95}
Another key ingredient here is the presence of a gapless excitation in the long wave-length limit,
as first proved by Hugenholtz and Pines. \cite{HP59}
This branch, which corresponds to the Bogoliubov mode in the weak-coupling regime, \cite{Bogoliubov47} 
can be identified now as the Nambu-Goldstone boson \cite{Goldstone61} of the spontaneously broken U(1) symmetry. 
The importance of the two features, i.e., ``conserving'' and ``gapless,''  for describing interacting condensed bosons 
was already pointed out by Hohenberg and Martin \cite{HM65}  in 1965 and also emphasized by Griffin \cite{Griffin96} 
soon after the realization of the Bose-Einstein condensation in the trapped atomic gases.
Despite considerable efforts, however, few systematic approximation schemes for BEC have been known to date which satisfy the two fundamental properties simultaneously. 
 
A notable exception may be the dielectric formalism. \cite{MW67,SK74,WG74,Griffin93}
It is designed specifically to describe another important feature of interacting condensed bosons that 
the single-particle spectrum and the two-particle density spectrum coincide, as first shown by Gavoret and Nozi\`ere. \cite{GN64}
By incorporating local number conservation additionally, 
it has provided the gapless spectrum of a weakly interacting homogeneous Bose gas beyond the leading order.\cite{WG74}  
However, a generalization of the formalism to inhomogeneous or nonequilibrium situations seems 
not straightforward when experiments on atomic gases are carried out with trap potentials dynamically. \cite{PS08,PS03,GNZ09}

Now, the main purpose of the present paper is 
to develop a self-consistent perturbation expansion for BEC
satisfying Goldstone's theorem \cite{Goldstone61} and dynamical conservation laws simultaneously.
This will be carried out by extending the normal-state Luttinger-Ward functional \cite{LW60}
so as to obey a couple of exact relations for BEC, i.e.,
that for the interaction energy and the Hugenholtz-Pines relation.
The two relations will be shown to yield a unique identical result for the functional $\Phi$ of BEC at least up to the third order in the self-consistent perturbation expansion.
The fact indicates that the series becomes exact when infinite terms are retained. 
It also turns out that the expansion reproduces the conserving-gapless mean field theory developed earlier with a subtraction procedure \cite{Kita05,Kita06} 
as the lowest-order approximation.
The formulation will be carried out in the coordinate space so that it is applicable to inhomogeneous systems
such as those under trap potentials and with vortices.
Using  the Keldysh Green's functions,
we will also extend it to describe nonequilibrium behaviors.
Those are subjects with many unresolved issues \cite{GNZ09} which cannot be treated 
by other theoretical methods for BEC 
such as the variational approach \cite{Feynman54,Feenberg69} and the Monte-Carlo method. \cite{Ceperley95}
The whole contents here are relevant to single-particle properties, and
we are planning to discuss two-particle properties 
in the near future.

The formalism will find a wide range of applications on BEC.
They include: (i) clarifying molecular field effects in condensed Bose systems
corresponding to the Landau theory of Fermi liquids;
(ii) nonequilibrium phenomena of BEC such as thermalization with full account of 
the quasiparticle collisions;
(iii) microscopic derivation of Landau two-fluid equations 
with definite interaction and temperature dependences of the
viscosity coefficient, etc.\
It will also be helpful to construct a practical functional for BEC within the density-functional formalism,
which still seems absent.

It is worth pointing out finally that the Luttinger-Ward functional is known as the ``two-particle irreducible (2PI) action'' in the relativistic quantum-field theory,\cite{CJT74,Berges04} and the difficulties mentioned above are also encountered in describing its broken symmetry phases such as that of the $\phi^{4}$ theory. \cite{BG77,IRK05} 
Thus, the present issue is relevant to a wide range of theoretical physics beyond BEC.

This paper is organized as follows. Section \ref{sec:exact_relations} summarizes exact results on an interacting Bose system including an expression of the thermodynamic potential for BEC with $\Phi$. Section \ref{sec:Phi} presents a definite procedure to construct $\Phi$. It is subsequently used to obtain a first few series of the self-consistent perturbation expansion explicitly.
Section \ref{sec:Eq-Prop} derives formally exact expressions of entropy and superfluid density in terms of Green's function which may also be useful for their approximate evaluations.
Section \ref{sec:real-time} extends the formulation to describe nonequilibrium dynamical evolutions of BEC.
Section \ref{sec:summary} summarizes the paper.
We put $\hbar=k_{\rm B}=1$ throughout with $k_{\rm B}$ the Boltzmann constant.

\section{Exact results\label{sec:exact_relations}}

We consider identical Bose particles with mass $m$ and spin $0$ described by the Hamiltonian:
\begin{equation}
H=H_{0}+H_{\rm int},
\label{Hamil}
\end{equation}
with
\begin{subequations}
\label{Hamil2}
\begin{equation}
H_{0}=\int d^{3}r_{1} \psi^{\dagger}({\bf r}_{1})K_{1}\psi({\bf r}_{1}),
\label{H_0}
\end{equation}
\begin{equation}
H_{\rm int}=\frac{1}{2}\int d^{3}r\int d^{3}r'\,\psi^{\dagger}({\bf r})
\psi^{\dagger}({\bf r}') V({\bf r}-{\bf r}')
 \psi({\bf r}')\psi({\bf r}) .
\label{H_int}
\end{equation}
\end{subequations}
Here $\psi^{\dagger}$ and $\psi$ are field operators satisfying the Bose commutation relations, 
${K}_{1}\equiv -\frac{1}{2m}\nabla_{1}^{2}-\mu$ with $\mu$  as the chemical potential, 
and $V$ is the interaction potential with the property $V({\bf r}-{\bf r}')=V({\bf r}'-{\bf r})$. 
Though dropped here, the effect of the trap potential can be included easily in $K_{1}$.

Let us introduce the Heisenberg representation of the field operators by  \cite{AGD63}
\begin{equation}
\psi(1) \equiv e^{\tau_{1}H}\psi({\bf r}_{1})e^{-\tau_{1}H} ,\hspace{5mm}
\bar{\psi}(1) \equiv e^{\tau_{1}H}\psi^{\,\dagger}({\bf r}_{1})e^{-\tau_{1}H}
,
\label{psi(1)-def}
\end{equation}
with $1\equiv ({\bf r}_{1},\tau_{1})$,
where $0 \leq \tau_{1} \leq T^{-1}$ with 
$T$ the temperature.
We next express $\psi(1)$
as a sum of the condensate wave function $\Psi(1)\equiv \langle
\psi(1)\rangle$ and the quasiparticle field $\phi(1)$
as 
\begin{equation}
\psi(1)=\Psi(1)+\phi(1),\hspace{5mm}
\bar{\psi}(1)=\bar{\Psi}(1)+\bar{\phi}(1),
\label{psi=Psi+psi-t}
\end{equation}
with $\langle\cdots\rangle$ denoting the grand-canonical average in terms of $H$. Note $\langle\phi(1)\rangle=0$ by definition.

Our Matsubara Green's function in the $2\times 2$ Nambu space is defined in terms of $\phi$ and $\bar{\phi}$ by
\begin{eqnarray}
&&\hspace{-10mm}
\hat{G}(1,2)\equiv - \left< T_{\tau}\!
\left[
\begin{array}{c}
\vspace{1mm}
\phi (1)\\ \bar{\phi}(1)
\end{array}
\right] \! [\, \bar{\phi}(2) \,\, \phi(2)\,] \right> 
\hat{\sigma}_{3}
\nonumber \\
&&\hspace{1.5mm}
\equiv \left[
\begin{array}{cc}
\vspace{1mm}
G(1,2) & F(1,2)
\\
-\bar{F}(1,2) & -\bar{G}(1,2)
\end{array}\right] ,
\label{hatG}
\end{eqnarray}
where $T_{\tau}$ denotes the ``time''-ordering operator \cite{AGD63}
and $\hat{\sigma}_{3}$ is the third Pauli matrix.
Every $2\times 2$ matrix in the Nambu space will be distinguished with the symbol $\hat{\,\,\,}$ 
on top of it like $\hat{G}$. 
The off-diagonal elements $F(1,2)=\langle T_{\tau}
\phi(1)\phi(2)\rangle$ and $\bar{F}(1,2)=\langle T_{\tau}
\bar{\phi}(1)\bar{\phi}(2)\rangle$ were introduced by Beliaev, \cite{Beliaev58} 
which describe the pair annihilation and creation of quasiparticles inherent in BEC.
The factor $\hat{\sigma}_{3}$ in Eq.\ (\ref{hatG})
is usually absent in the definition of
$\hat{G}$; \cite{HM65,Griffin96}
it brings an advantage that 
poles of $\hat{G}$ directly correspond to the
Bogoliubov quasiparticles. \cite{Kita05,Kita06}

It is easily checked that the elements of $\hat{G}$ satisfy $G^{*}(1,2)=G({\bf r}_{2}\tau_{1},{\bf r}_{1}\tau_{2})$, $F(1,2)=F(2,1)$, 
$\bar{G}(1,2)=G(2,1)$, and $\bar{F}(1,2)=F^{*}({\bf r}_{2}\tau_{1},{\bf r}_{1}\tau_{2})$,
with superscript $^{*}$ denoting complex conjugate.
These four relations are expressed compactly in terms of $\hat{G}$ as
\begin{subequations}
\label{hatG-symm}
\begin{eqnarray}
&&
\hat{\sigma}_{3}\hat{G}^{\rm \dagger}(1,2)\hat{\sigma}_{3}=\hat{G}({\bf r}_{2}\tau_{1},{\bf r}_{1}\tau_{2}),
\\
&&
\hat{\sigma}_{1}\hat{G}^{*}(1,2)\hat{\sigma}_{1}=-\hat{G}({\bf r}_{1}\tau_{2},{\bf r}_{2}\tau_{1}),
\end{eqnarray}
\end{subequations}
where superscript $^{\rm \dagger}$ denotes Hermitian conjugate in the matrix algebra.
Using $\hat{G}^{-1}\hat{G}=\hat{1}$, one can show that $\hat{G}^{-1}$ also obeys the relations of Eq.\ (\ref{hatG-symm}).

Total particle number $N$ is calculated by integrating $\langle\bar{\psi}(1)\psi(1)\rangle$
over the whole space of the system, i.e.,
\begin{equation}
N=\int d^3 r_{1} \bigl[ \bar{\Psi}(1)\Psi(1)-G(1,1_{+})\bigr] ,
\label{N}
\end{equation}
where the subscript of $1_{+}$ denotes an extra infinitesimal positive constant in the argument 
$\tau_{1}$ to put the creation operator to the left for the equal-time average. \cite{LW60}
Equation (\ref{N}) may also be used to eliminate $\mu$ in favor of $N$.
To be explicit, we will proceed by choosing $(T,\mu)$ as independent variables,
which will be dropped in most cases.

We summarize relevant exact results on the system below.
Those of Secs.\ \ref{subsec:DBeq} and \ref{subsec:HP-th} 
have been known from the late 1950s.
However, they often have been proved or reviewed only for uniform systems 
with different notations.
We hence provide in Appendix A detailed derivations of those results
together with that of Sec.\ \ref{subsec:IntE}
for general inhomogeneous systems
so as to be compatible with definition (\ref{hatG}) of our Green's function.

\subsection{Dyson-Beliaev equation\label{subsec:DBeq}}

Equation (\ref{hatG}) satisfies the Dyson-Beliaev equation: \cite{Beliaev58,HP59}
\begin{equation}
\int d3 \bigl[\hat{G}_{0}^{-1}(1,3)-\hat{\Sigma}(1,3)\bigr]\hat{G}(3,2)
=\hat{\sigma}_{0}\delta(1,2) ,
\label{Dyson}
\end{equation}
where $\hat{G}_{0}^{-1}$ is defined by
\begin{equation}
\hat{G}_{0}^{-1}(1,2)\equiv\biggl(-\hat{\sigma}_{0}\frac{\partial}{\partial \tau_{1}}-\hat{\sigma}_{3}K_{1}\biggr)\delta(1,2),
\label{hatG_0^-1}
\end{equation}
with $\hat{\sigma}_{0}$ as the $2\times 2$ unit matrix, $\hat{\Sigma}$ the self-energy, and
$\delta(1,2)\equiv \delta(\tau_{1}-\tau_{2})\delta({\bf r}_{1}-{\bf r}_{2})$.
A proof of Eq.\ (\ref{Dyson}) is given in Appendix \ref{subsec:DB-proof}.

It follows from $\hat{\Sigma}=\hat{G}_{0}^{-1}-\hat{G}^{-1}$ and the symmetry of $\hat{G}^{-1}$ that
$\hat{\Sigma}$ also obeys the relations of Eq.\ (\ref{hatG-symm}).
Let us write it as
\begin{equation}
\hat{\Sigma}(1,2)= \left[
\begin{array}{cc}
\vspace{1mm}
\Sigma(1,2) & \Delta(1,2)
\\
-\bar{\Delta}(1,2) & -\bar{\Sigma}(1,2)
\end{array}\right] .
\label{hatSigma}
\end{equation}
It then follows that the elements satisfy 
$\Sigma^{*}(1,2)=\Sigma({\bf r}_{2}\tau_{1},{\bf r}_{1}\tau_{2})$, $\Delta(1,2)=\Delta(2,1)$, 
$\bar{\Sigma}(1,2)=\Sigma(2,1)$, and $\bar{\Delta}(1,2)=\Delta^{*}({\bf r}_{2}\tau_{1},{\bf r}_{1}\tau_{2})$.

\subsection{Hugenholtz-Pines theorem\label{subsec:HP-th}}

As shown in Appendix \ref{subsec:HP-proof}, the Hugenholtz-Pines theorem \cite{HP59}
can be extended to inhomogeneous systems as
\begin{equation}
\int d2 \bigl[  \hat{G}_{0}^{-1}(1,2)-\hat{\Sigma}(1,2)\bigr]
\left[\begin{array}{c}
\vspace{1mm}
\Psi(2) \\ -\bar{\Psi}(2)
\end{array}\right]=\left[\begin{array}{c}
0\\ 0
\end{array}\right] .
\label{GP}
\end{equation}
This is the condition for the excitation spectra to have a gapless mode as compatible with Goldstone's theorem. \cite{PS95,Goldstone61}
In the homogeneous case of $\Psi(1)=\sqrt{n_{0}}$ with $n_{0}$ as the condensate density, the first row of Eq.\ (\ref{GP}) in the Fourier space reduces to the familiar Hugenholtz-Pines relation: \cite{HP59} $\mu=\Sigma_{p=0}-\Delta_{p=0}$,
where $p\equiv ({\bf p},i\varepsilon_{n})$ is the four momentum with
$\varepsilon_{n}\equiv 2n\pi T$ as the Matsubara frequency ($n=0,\pm 1,\pm 2,\cdots$).
Note that Eq.\ (\ref{GP}) is given here in terms of the same operator as the Dyson-Beliaev equation (\ref{Dyson}); 
see the comment below Eq.\ (\ref{hatu}) for its relevance to Goldstone's theorem.
Equation (\ref{GP}) can also be regarded as the generalized 
Gross-Pitaevskii equation \cite{Gross61,Pitaevskii61} to incorporate the quasiparticle contribution 
into the self-energies.

It is worth pointing out that our proof of Eq.\ (\ref{GP}) in Appendix \ref{subsec:HP-proof}
has been carried out by using the gauge transformation
relevant to the broken U(1) symmetry without making any specific assumptions on the structure of
the self-energies.
Especially, it has removed the implicit supposition by Hugenholtz and Pines \cite{HP59} that 
the self-energies in condensed Bose systems also be proper in the conventional sense. \cite{LW60}

\subsection{Interaction energy\label{subsec:IntE}}

It is shown in Appendix \ref{subsec:IntE-proof} 
that the interaction energy $\langle H_{\rm int}\rangle$, i.e., the grand-canonical average of Eq.\ (\ref{H_int}),
can be expressed in terms of the self-energies as
\begin{eqnarray}
&&\hspace{-5mm}
\langle H_{\rm int}\rangle 
=\frac{T}{4}\int d1\int d2 \,\bigl\{ 2\Sigma(1,2) [\Psi(2)\bar{\Psi}(1)- G(2,1_{+})]
\nonumber \\
&&\hspace{9mm}
-\bar{\Delta}(1,2) [\Psi(2)\Psi(1)- F(2,1)] 
\nonumber \\
&&\hspace{9mm}
-\Delta(1,2) [\bar{\Psi}(2)\bar{\Psi}(1)-\bar{F}(2,1)]\bigr\}.
\label{<H_int>}
\end{eqnarray}
This is one of the key relations indispensable below.

\subsection{Luttinger-Ward functional}

Luttinger and Ward \cite{LW60} gave an expression of the thermodynamic potential $\Omega\equiv -T\ln {\rm Tr}\,e^{- H/T}$ for an interacting normal Fermi system as a functional of the self-energy $\Sigma$. Their consideration can be extended easily to the normal Bose system of  $\Psi=0$. It is more convenient to regard the resultant $\Omega$ as a functional of $G$, which reads as
\begin{equation}
\Omega  = T\,{\rm Tr}\bigl[\ln (-G_{0}^{-1}+\Sigma)+\Sigma G\bigr] +\Phi,
\label{Omega_LW}
\end{equation}
with ${\rm Tr}AB\equiv\int d1\int d2 A(1,2)B(2,1_{+})$ and
\begin{equation}
G_{0}^{-1}(1,2)\equiv \biggl(- \frac{\partial}{\partial \tau_{1}}-K_{1}\biggr)\delta(1,2) .
\label{G_0^-1}
\end{equation}
The quantity $\Phi$ denotes contribution of all the skeleton diagrams in the simple perturbation expansion for $\Omega$ 
with the replacement $G_{0}\rightarrow G$. \cite{LW60} Its functional derivative with respect to $G$ yields the self-energy $\Sigma$ as 
\begin{equation}
\Sigma(1,2)=-T^{-1} \frac{\delta \Phi}{\delta G(2,1)}.
\label{Sigma-G}
\end{equation}
It hence follows from Dyson's equation $G=(G_{0}^{-1}-\Sigma)^{-1}$ that $\Omega$ is stationary with respect to 
$G$ as ${\delta \Omega}/{\delta G(2,1)}=0$.

Luttinger and Ward also put Eq.\ (\ref{Sigma-G}) into an integral form with the $n$th-order self-energy $\Sigma^{(n)}$ in terms of the interaction.
To be explicit, $\Sigma^{(n)}$ is defined as the contribution of all the $n$th-order skeleton diagrams 
in the simple perturbation expansion for the proper self-energy with the replacement 
$G_{0}\rightarrow G$. \cite{LW60}
Noting that there are $2n-1$ Green's-function lines in the diagrams of $\Sigma^{(n)}$,
Eq.\ (\ref{Sigma-G}) can be integrated order by order into \cite{LW60}
\begin{equation}
\Phi =-T\sum_{n=1}^{\infty}\frac{1}{2n}{\rm Tr}\Sigma^{(n)} G.
\label{Phi-normal}
\end{equation}
Comparing this expression with Eq.\ (\ref{<H_int>}) of the normal state ($\Psi=0$, $F=0$),
we obtain a relation between the $n$th-order terms as
\begin{equation}
\Phi^{(n)} =\frac{1}{n}\langle H_{\rm int}\rangle^{(n)}.
\label{Phi-H_int}
\end{equation}
The factor $1/n$ is due to the extra $H_{\rm int}$ present in the evaluation of $\langle H_{\rm int}\rangle^{(n)}$ compared with that of $\Phi^{(n)}$.
Hence the relation will hold true generally in self-consistent perturbation expansions beyond the normal phase.

\subsection{De Dominicis-Martin theorem\label{subsec:DM-th}}

Using a series of Legendre transformations, it was shown by de Dominicis and Martin \cite{dDM64} 
(see also Refs.\ \onlinecite{CJT74} and \onlinecite{Berges04}) that the thermodynamic potential $\Omega$ in the condensed phase can be expressed as a functional $\Omega[G,F,\bar{F},\Psi,\bar{\Psi}]$ such that
\begin{eqnarray}
\frac{\delta\Omega}{\delta G(2,1)}=\frac{\delta\Omega}{\delta \bar{F}(2,1)}=0, \hspace{5mm} 
\frac{\delta\Omega}{\delta \bar{\Psi}(1)}=0.
\label{stationarity}
\end{eqnarray}
Thus, the exact thermodynamic potential is stationary with respect to variations in both 
the condensate wave function and Green's functions.
Equation (\ref{stationarity}) generalizes ${\delta \Omega}/{\delta G(2,1)}=0$ of the normal-state Luttinger-Ward functional [Eq.\ (\ref{Omega_LW})] to condensed Bose systems. However, no explicit $\Omega$ has been known for BEC which satisfies Eq.\ (\ref{stationarity}) via Eqs.\ (\ref{Dyson}) and (\ref{GP}) and also includes Eq.\ (\ref{Omega_LW}) as the limit $\Psi\rightarrow 0$.

Following Eq.\ (\ref{Omega_LW}) for the normal state, 
we now express $\Omega$ of the condensed phase as
\begin{eqnarray}
&&\hspace{-10mm}
\Omega =-T\int d1\int d2 \,\bar{\Psi}(1)G_{0}^{-1}(1,2)\Psi(2)
\nonumber \\
&&\hspace{-2.5mm}
+
 \frac{T}{2}{\rm Tr}\bigl[\ln (-\hat{G}_{0}^{-1}+\hat{\Sigma})+\hat{\Sigma} \hat{G}\bigr] +\Phi,
\label{Omega}
\end{eqnarray}
where $G_{0}^{-1}$ and $\hat{G}_{0}^{-1}$ are given as Eqs.\ (\ref{G_0^-1}) and 
(\ref{hatG_0^-1}), respectively, and ${\rm Tr}$ is now defined by
\begin{eqnarray}
&&\hspace{-10mm}
{\rm Tr}\,\hat{\Sigma} \hat{G}\equiv \int d1\int d2\,
 {\rm Tr}\left[
\begin{array}{cc}
\vspace{1mm}
\Sigma(1,2) & \Delta(1,2)
\\
-\bar{\Delta}(1,2) & -\bar{\Sigma}(1,2)
\end{array}\right] 
\nonumber \\
&&\hspace{4mm}\times
 \left[
\begin{array}{cc}
\vspace{1mm}
G(2,1_{+}) & F(2,1)
\\
-\bar{F}(2,1) & -\bar{G}(2,1_{-})
\end{array}\right] .
\label{Tr-def}
\end{eqnarray}
The subscript of $1_{-}$ denotes an extra infinitesimal negative constant in $\tau_{1}$  to 
put the creation operator to the left
for equal-time averages, and the second Tr denotes the usual 
trace in the matrix algebra.
Equation (\ref{Omega}) appropriately reduces to Eq.\ (\ref{Omega_LW}) as $\Psi\rightarrow 0$.
Whereas the first two terms in Eq.\ (\ref{Omega}) remain finite even for the ideal Bose gas, 
$\Phi$ is made up only of contribution due to the interaction.
This is one of the advantages for adopting expression (\ref{Omega}).

Once $\Omega$ is written as Eq.\  (\ref{Omega}),
one can show by using Eq.\ (\ref{Dyson}), Eq.\ (\ref{GP}), $\bar{G}(1,2)=G(2,1)$, and  
$\bar{\Sigma}(1,2)=\Sigma(2,1)$ that
condition (\ref{stationarity}) can be expressed equivalently with respect to $\Phi$
as
\begin{subequations}
\label{Phi-deriv}
\begin{equation}
\Sigma(1,2)=-T^{-1}\frac{\partial \Phi}{\partial G(2,1)}, \hspace{5mm}
\Delta(1,2)=2T^{-1} \frac{\partial \Phi}{\partial \bar{F}(2,1)}, 
\label{Sigma-Phi}
\end{equation}
\begin{equation}
T^{-1}\frac{\partial \Phi}{\partial \bar{\Psi}(1)}=\int d2\, \bigl[ \Sigma(1,2)\Psi(2)-\Delta(1,2)\bar{\Psi}(2)\bigr] . 
\label{Psi-Phi}
\end{equation}
\end{subequations}
These are direct generalizations of Eq.\ (\ref{Sigma-G}) for the normal state into the condensed phase.

Now, our gapless $\Phi$-derivable approximation denotes (i) constructing $\Phi$ so as to reproduce 
Eq.\ (\ref{Phi-deriv})
and (ii) determining $\hat{G}$, $\Psi$, and $\hat{\Sigma}$ self-consistently with Eqs.\ (\ref{Dyson}), (\ref{GP}),
and (\ref{Phi-deriv}). 
It will obey dynamical conservation laws \cite{Baym62,Kita06} as well as Goldstone's theorem, \cite{PS95,Goldstone61}
thereby enabling us to handle equilibrium and nonequilibrium phenomena of BEC on an equal footing
with the Nambu-Goldstone boson.

It is worth pointing out that expression (\ref{Omega}) becomes exact 
when $\Phi$ satisfies Eq.\ (\ref{Phi-H_int}) at each order up to $n=\infty$ in the self-consistent perturbation expansion.
With $H_{\rm int}\rightarrow \lambda H_{\rm int}$ in Eq.\ (\ref{Hamil}),
the proof proceeds in exactly the same way as that of the normal state \cite{LW60}
as follows.
First, we find with Eqs.\  (\ref{Phi-H_int}) and (\ref{stationarity})
that the corresponding expression of Eq.\ (\ref{Omega}) obeys
the same first order differential equation $\partial \Omega_{\lambda}/\partial \lambda=\langle \lambda H_{\rm int}\rangle_{\lambda}/\lambda$ as the 
defining one $\Omega_{\lambda}\equiv -T\ln {\rm Tr}\,e^{- (H_{0}+\lambda H_{\rm int})/T}$, where $\langle \cdots\rangle_{\lambda}$ denotes
the grand-canonical average in terms of $H_{0}+\lambda H_{\rm int}$.
Second, the two expressions yield the same initial value $\Omega_{\lambda=0}$.
We hence arrive at the above conclusion.
Thus, the gapless $\Phi$-derivable scheme obeying Eq.\ (\ref{Phi-H_int}) also includes the exact theory as a limit.

\subsection{Exact relations with $\Phi$}

We now present a couple of exact relations in terms of $\Phi$ to be satisfied in the condensed phase.
Let us substitute the $n$th-order contribution of Eq.\ (\ref{<H_int>}) into Eq.\ (\ref{Phi-H_int}) and subsequently use Eq.\ (\ref{Sigma-Phi}).
We thereby obtain
\begin{eqnarray}
&&\hspace{-5mm}
2n \Phi^{(n)} + \int d1\int d2 \biggl\{\frac{\delta \Phi^{(n)}}{\delta G(2,1)} [\Psi(2)\bar{\Psi}(1)-G(2,1)]
\nonumber \\
&&\hspace{-5mm}
+\frac{\delta \Phi^{(n)}}{\delta F(2,1)} [\Psi(2)\Psi(1)-F(2,1)] 
\nonumber \\
&&\hspace{-5mm}
+\frac{\delta \Phi^{(n)}}{\delta \bar{F}(2,1)} [\bar{\Psi}(2)\bar{\Psi}(1)
-\bar{F}(2,1)]\biggr\}=0.
\label{relation1}
\end{eqnarray}
Next, substitution of  Eq.\ (\ref{Sigma-Phi}) into Eq.\ (\ref{Psi-Phi}) yields
\begin{equation}
\frac{\delta \Phi}{\delta \bar{\Psi}(1)}
+\int d2\biggl[\frac{\delta \Phi}{\delta G(2,1)}\Psi(2)+2\frac{\delta \Phi}{\delta \bar{F}(2,1)} \bar{\Psi}(2)\biggr]=0 .
\label{relation2}
\end{equation}
The above two equalities will play a crucial role below for writing $\Phi$ down explicitly.

\section{Constructing $\Phi$\label{sec:Phi}}
 
One of the basic difficulties in developing the self-consistent perturbation expansion for BEC may be attributed to the absence of a definite concept of ``skeleton diagrams,''
which were clear in normal systems, \cite{LW60}
due to the appearance of finite one-particle average $\Psi(1)\equiv \langle\psi(1)\rangle$.
It brings an ambiguity as to how to count the contribution with $F$ and $\Psi$ adequately in the renormalization process.
Our approach here is to determine the contribution to $\Phi$  inherent in BEC with the exact relation of Eq.\ (\ref{relation1}) or Eq.\ (\ref{relation2}) so as to reproduce the Luttinger-Ward functional for $\Psi\rightarrow 0$, 
thereby avoiding the conceptual difficulty to define ``skeleton diagrams'' explicitly.
The two relations will be shown to yield a unique
result at each order in the self-consistent perturbation expansion in terms of the interaction.

To this end, we introduce the symmetrized vertex: \cite{AGD63}
\begin{eqnarray}
&&\hspace{-9mm}
\Gamma^{(0)}(11',22')
\equiv V({\bf r}_{1}-{\bf r}_{2})\delta(\tau_{1}-\tau_{2})
\nonumber \\
&&\hspace{14.3mm}\times
[\delta(1,1')\delta(2,2')+\delta(1,2')\delta(2,1')] ,
\label{Gamma^(0)}
\end{eqnarray}
satisfying $\Gamma^{(0)}(11',22')=\Gamma^{(0)}(22',11')=\Gamma^{(0)}(1'1,2'2)=\Gamma^{(0)}(12',21')$.
It helps us to reduce relevant Feynman diagrams substantially at the expense of introducing 
some cumbersomeness in the calculation of numerical factors. \cite{AGD63}
Our consideration below will be carried out in terms of topologically distinct diagrams,
where $G$, $F$, and $\bar{F}$ are expressed in the same way as those of superconductivity, \cite{AGD63}  and $\Gamma^{(0)}$ is denoted by a filled circle.
Following Popov, \cite{Popov87} we also suppress drawing symbols for $\Psi$ and $\bar{\Psi}$ in those diagrams;
they can be recovered easily with the fact that Eq.\ (\ref{H_int}) originally contains two pairs of creation and annihilation operators.

\begin{figure}[b]
        \begin{center}
                \includegraphics[width=0.8\linewidth]{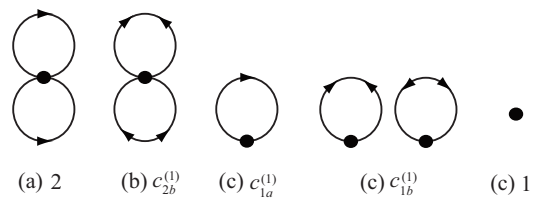}
        \end{center}
        \caption{Diagrams contributing to $\Phi^{(1)}$. Here (a)-(c) distinguish three kinds of diagrams considered at different stages of the procedure in Sec.\ \ref{subsec:Proc-Phi},
and numbers and unknown variables $c^{(1)}_{\nu}$ ($\nu=2b,1a,1b$) denote relative weights of those diagrams.  
Each coefficient should be multiplied by $T/4$ to 
        obtain the absolute weight.}
        \label{fig:Phi^(1)}
\end{figure}

\subsection{Definite procedure for $\Phi^{(n)}$\label{subsec:Proc-Phi}}

Consider the $n$th-order contribution. 
The procedure to construct $\Phi^{(n)}$ is summarized as (a)-(d) below.
See Figs.\ 1 and 2 as explicit examples of relevant diagrams for  $n=1$ and $2$, respectively.

(a) Draw all the normal-state diagrams contributing to $\Phi^{(n)}$, i.e., those diagrams which appear in the Luttinger-Ward functional. \cite{LW60}
With each such diagram, associate the factor of the normal state.

(b) Draw all the distinct diagrams obtained from those of (a) 
by successively changing the directions of a pair of incoming and outgoing arrows at each vertex.
This enumerates all the processes where $F$ or $\bar{F}$ is relevant in place of $G$.
With each such diagram,  associate an unknown coefficient.

(c1) Draw all the distinct diagrams obtained from those of (a) and (b) by successively removing a Green's-function line so as to meet the condition that a further removal of any line from each diagram would not break it into two unconnected parts. 
The procedure incorporates all the processes where the condensate wave function participates explicitly.
The latter condition guarantees that the self-energies obtained by Eq.\ (\ref{Sigma-Phi}) are composed of connected diagrams.

(c2) 
Associate an unknown coefficient with each such diagram, except the one consisting only of a single vertex in the first order, i.e., the rightmost diagram in Fig.\ 1,
for which the coefficient is easily identified to be $T/4$.  Indeed, the latter represents the term obtained from Eq.\ (\ref{H_int}) by replacing every field operator by its expectation value, i.e., the condensate wave function.

(d) Determine the unknown coefficients of (b) and (c) by requiring that either Eq.\ (\ref{relation1}) or Eq.\ (\ref{relation2}) be satisfied.

It is worth pointing out that, with Eq.\ (\ref{Sigma-Phi}), the diagrams of (c1) necessarily yields those self-energies which are separated into two  parts by cutting a single line, i.e.,
those classified as ``improper'' in the conventional sense. \cite{LW60}
According to Eq.\ (\ref{<H_int>}), however, we surely need to consider this kind of self-energy diagrams 
in the self-consistent perturbation expansion for BEC.
It should also be mentioned that our proof of Eq.\ (\ref{GP}) in Appendix \ref{subsec:HP-proof}
is carried out without the implicit supposition by Hugenholtz and Pines \cite{HP59} 
that the self-energies be ``proper''
in the conventional sense. \cite{LW60}
Thus, there is nothing inconsistent on this point in our formulation.

\begin{figure}[t]
        \begin{center}
                \includegraphics[width=0.9\linewidth]{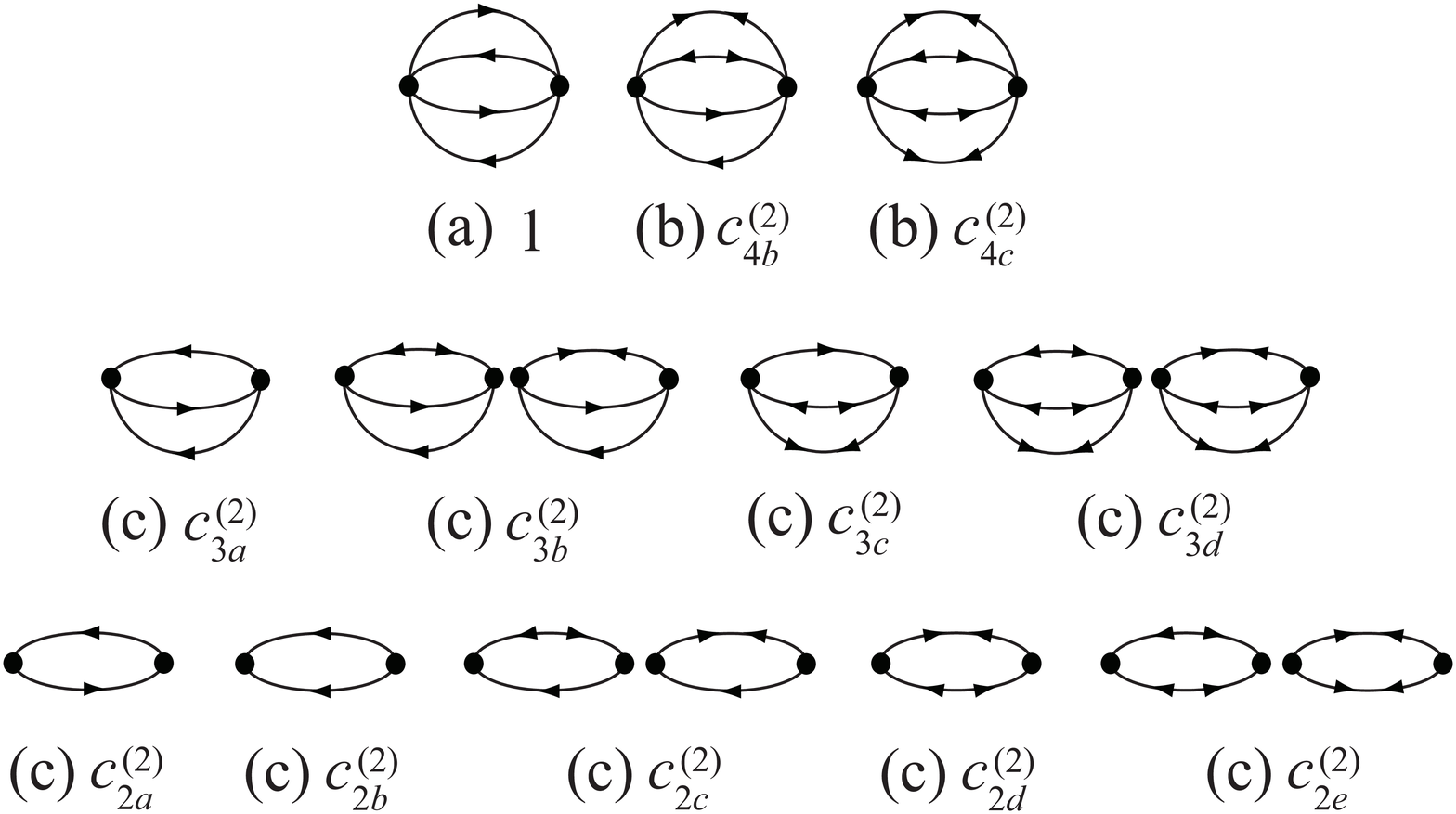}
        \end{center}
        \caption{Diagrams contributing to $\Phi^{(2)}$. Here (a)-(c) distinguish three kinds of diagrams considered at different stages of the procedure in Sec.\ \ref{subsec:Proc-Phi}, 
        and the number $1$ and unknown variables $c^{(2)}_{\nu}$ ($\nu=4b,\cdots,2e$) denote relative weights of those diagrams.  Each coefficient should be multiplied by $-T/8$ to obtain the absolute weight.}
        \label{fig:Phi^(2)}
\end{figure}

\subsection{Expression of $\Phi^{(1)}$}

The first-order contribution to $\Phi$ is given by the diagrams of Fig.\ \ref{fig:Phi^(1)}. 
They can be expressed analytically as
\begin{eqnarray}
&&\hspace{-4mm}
\Phi^{(1)}=\frac{T}{4} \int d1 \int d1' \int d2 \int d2'\,\Gamma^{(0)}(11',22')
\nonumber \\
&&\hspace{7mm}
\times \bigl\{ 2G(1,1')G(2,2')
+c^{(1)}_{2b}F(1,2)\bar{F}(1',2')
\nonumber \\
&&\hspace{7mm}
+c^{(1)}_{1a}G(1,1')\Psi(2)\bar{\Psi}(2')
+c^{(1)}_{1b}\bigl[F(1,2)\bar{\Psi}(1')\bar{\Psi}(2')
\nonumber \\
&&\hspace{7mm}
+\bar{F}(1',2')\Psi(1)\Psi(2)\bigr]
+\bar{\Psi}(1')\bar{\Psi}(2')\Psi(2)\Psi(1) \bigr\} ,
\nonumber \\
\label{Phi^(1)}
\end{eqnarray}
where $c^{(1)}_{2b}$, $c^{(1)}_{1a}$, and $c^{(1)}_{1b}$ are unknown coefficients.

We now require that Eq.\ (\ref{relation1}) be satisfied. 
It turns out that contribution of the first two diagrams in Fig.\ 1 vanishes in the equation. 
This cancellation is characteristic of those diagrams with no condensate wave function
and holds order by order in Eq.\ (\ref{relation1}). 
Hence relevant to Eq.\  (\ref{relation1}) in Fig.\ 1 are the last three diagrams, 
which yield
$$
4+c^{(1)}_{1a}=0,\hspace{5mm}c^{(1)}_{2b}+c^{(1)}_{1b}=0,\hspace{5mm}c^{(1)}_{1a}+2c^{(1)}_{1b}+2=0 ,
$$
respectively.
We hence obtain
\begin{equation}
c^{(1)}_{2b}=-1,\hspace{5mm}c^{(1)}_{1a}=-4,\hspace{5mm}c^{(1)}_{1b}=1.
\label{c^(1)}
\end{equation}
Thus, the coefficients $c^{(1)}_{\nu}$ have been determined uniquely.

Alternatively, we impose Eq.\ (\ref{relation2}).
Terms in the resultant equation can be expressed graphically by the last three diagrams of Fig.\ 1
with an extra symbol for the missing $\bar{\Psi}$ at each vertex.
We obtain the same equations as above in terms of $c^{(1)}_{2b}$, $c^{(1)}_{1a}$, and $c^{(1)}_{1b}$. 
We hence arrive at Eq.\ (\ref{c^(1)}) again.

It is worth pointing out that, 
except for the signs of $F$ and $\bar{F}$ in the definition of Eq.\ (\ref{hatG}),
functional (\ref{Phi^(1)}) with coefficient  (\ref{c^(1)}) 
is exactly identical to that of the conserving-gapless mean-field theory \cite{Kita05,Kita06} developed earlier with a subtraction procedure. 
Moreover, those signs have been shown not to affect the physical quantities at all within the first order. \cite{Kita06}
Thus, the mean-field theory has been identified here as the only self-consistent theory of the first order  compatible with exact relations (\ref{relation1}) and (\ref{relation2}).

\subsection{Expression of $\Phi^{(2)}$}

The second-order contribution to $\Phi$ is  given by the diagrams of Fig.\ \ref{fig:Phi^(2)}. 
They are expressed analytically as
\begin{widetext}
\begin{eqnarray}
&&\hspace{-5mm}
\Phi^{(2)}
=-\frac{T}{8} \int d1 \cdots \int d4'\,
\Gamma^{(0)}(11',22')\Gamma^{(0)}(33',44')
\bigl\{G(2,3')G(3,2')G(1,4')G(4,1')
\nonumber \\
&&\hspace{6mm}
+c^{(2)}_{4b}\bar{F}(2',3')F(2,3)G(1,4')G(4,1')
+c^{(2)}_{4c}\bar{F}(2',3')F(2,3)\bar{F}(1',4')F(1,4)
\nonumber \\
&&\hspace{6mm}
+c^{(2)}_{3a}G(2,3')G(3,2')G(1,4')\Psi(4)\bar{\Psi}(1')
+c^{(2)}_{3b}\bigl[\bar{F}(2',3')\Psi(2)\Psi(3)
+F(2,3)\bar{\Psi}(2')\bar{\Psi}(3')\bigr]
G(1,4')G(4,1')
\nonumber \\
&&\hspace{6mm}
+c^{(2)}_{3c}\bar{F}(2',3')F(2,3)G(1,4')\Psi(4)\bar{\Psi}(1')
+c^{(2)}_{3d}\bigl[\bar{F}(2',3')\Psi(2)\Psi(3)+
F(2,3)\bar{\Psi}(2')\bar{\Psi}(3')\bigr]
\bar{F}(1',4')F(1,4)
\nonumber \\
&&\hspace{6mm}
+c^{(2)}_{2a}G(2,3')G(3,2')\Psi(1)\bar{\Psi}(4')\Psi(4)\bar{\Psi}(1')
+c^{(2)}_{2b}G(2,3')\Psi(3)\bar{\Psi}(2')G(1,4')\Psi(4)\bar{\Psi}(1')
\nonumber \\
&&\hspace{6mm}
+c^{(2)}_{2c}\bigl[\bar{F}(2',3')\Psi(2)\Psi(3)
+F(2,3)\bar{\Psi}(2')\bar{\Psi}(3')\bigr]
G(1,4')\Psi(4)\bar{\Psi}(1')
+c^{(2)}_{2d}\bar{F}(2',3')F(2,3)\Psi(1)\bar{\Psi}(4')\Psi(4)\bar{\Psi}(1')
\nonumber \\
&&\hspace{6mm}
+c^{(2)}_{2e}\bigl[F(2,3)\bar{\Psi}(3')\bar{\Psi}(2')F(1,4)
\bar{\Psi}(4')\bar{\Psi}(1')
+\bar{F}(2',3')\Psi(3)\Psi(2)\bar{F}(1',4')
\Psi(4)\Psi(1)\bigr]
\bigr\} .
\label{Phi^(2)}
\end{eqnarray}
\end{widetext}
Here the first term in the curly brackets corresponds to the normal-state process, whereas the others  with 
unknown prefactors $c^{(2)}_{\nu}$ ($\nu=4b,\cdots,2e$) are characteristic of BEC.

\begin{figure}[t]
        \begin{center}
                \includegraphics[width=0.4\linewidth]{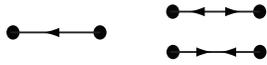}
        \end{center}
        \caption{Two kinds of extra diagrams which appear in the calculation of Eq.\ (\ref{relation1}) for $n=2$.}
        \label{fig:Phi^(2)-A}
\end{figure}

We now impose Eq.\ (\ref{relation1}) on the unknown coefficients.
It yields eleven algebraic equations originating from the prefactors of (i) the diagrams in the second and third rows of Fig.\ 2
and (ii) two kinds of diagrams in Fig.\ 3.
They are given by
\begin{subequations}
\label{c^(2)-eq}
\begin{equation}
0=c^{(2)}_{3a}+4=c^{(2)}_{3b}+ c^{(2)}_{4b}
=c^{(2)}_{3c}+2c^{(2)}_{4b}=c^{(2)}_{3d}+ 2c^{(2)}_{4c},
\label{c^(2)-eq1}
\end{equation}
\begin{eqnarray}
&&\hspace{-3mm}
0=c^{(2)}_{2a}+c^{(2)}_{3a}+c^{(2)}_{3b}=2c^{(2)}_{2b}+c^{(2)}_{3a}=2c^{(2)}_{2c}+c^{(2)}_{3c}+2c^{(2)}_{3b}
\nonumber \\
&&\hspace{-0.3mm} =
2c^{(2)}_{2d}+c^{(2)}_{3c}+4c^{(2)}_{3d}=2c^{(2)}_{2e}+c^{(2)}_{3d},
\label{c^(2)-eq2}
\end{eqnarray}
\begin{equation}
0=c^{(2)}_{2a}+c^{(2)}_{2b}+c^{(2)}_{2c}=
c^{(2)}_{2c}+c^{(2)}_{2d}+2c^{(2)}_{2e},
\label{c^(2)-eq3}
\end{equation}
\end{subequations}
respectively.
Solving them, we obtain
\begin{eqnarray} 
&& \hspace{15mm}
c^{(2)}_{4b}=-2, \hspace{3mm} c^{(2)}_{4c}=1,
\nonumber \\
&& \hspace{-4mm}
c^{(2)}_{3a}=-4,\hspace{3mm}c^{(2)}_{3b}=2,\hspace{3mm}  c^{(2)}_{3c}=4,\hspace{3mm}  
c^{(2)}_{3d}=-2,
\nonumber \\
&& \hspace{-8mm}
c^{(2)}_{2a}=c^{(2)}_{2b}=2,\hspace{3mm} c^{(2)}_{2c}=-4,\hspace{3mm} c^{(2)}_{2d}=2,\hspace{3mm} c^{(2)}_{2e}=1.
\label{c^(2)}
\end{eqnarray}
Thus, the coefficients $c^{(2)}_{\nu}$ have been determined uniquely with Eq.\ (\ref{relation1}).

\begin{figure}[b]
        \begin{center}
                \includegraphics[width=0.6\linewidth]{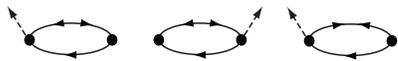}
        \end{center}
        \caption{Three distinct diagrams necessary in the calculation of Eq.\ (\ref{relation2}) for $n=2$. 
         They are obtained from the pair of diagrams with the coefficient $c^{(2)}_{2c}$ in Fig.\ 2 by adding a broken-line arrow 
         for the missing $\bar{\Psi}$ at a single vertex.}
        \label{fig:Phi^(2)-B}
\end{figure}

We may alternatively require that Eq.\ (\ref{relation2})  be satisfied. 
Terms which appear in the calculation of Eq.\ (\ref{relation2})  can also be expressed graphically.
They are obtained from the diagrams in the second and third row of Fig.\ 2
and those of Fig.\ 3 by adding a broken-line arrow for the missing $\bar{\Psi}$ at a vertex 
in all possible ways.
The insertion is topologically unique for most of the diagrams, e.g., those in the second row of Fig.\ 2; in this case the corresponding equation for each diagram turns out to be the same as that
in Eq.\ (\ref{c^(2)-eq}). There are three exceptions.
The first of them corresponds to the pair of diagrams with the coefficient $c_{2c}^{(2)}$ in Fig.\ 2,
which yields three distinct diagrams of Fig.\ 4.
Thus, $2c^{(2)}_{2c}+c^{(2)}_{3c}+2c^{(2)}_{3b}=0$ in Eq.\ (\ref{c^(2)-eq2}) is now replaced by the three equations:
\begin{subequations}
\label{c^(2)-eq-relation2}
\begin{equation}
0=c^{(2)}_{2c}+2c^{(2)}_{3b}=2c^{(2)}_{2c}+2c^{(2)}_{3b}+c^{(2)}_{3c}=c^{(2)}_{2c}+c^{(2)}_{3c} . 
\end{equation}
The others are two kinds of diagrams in Fig.\ 3, for which there are four different ways to add a broken-line arrow.
The corresponding equations are given by
\begin{equation}
0=2c^{(2)}_{2a}+c^{(2)}_{2c}=2c^{(2)}_{2b}+c^{(2)}_{2c}=c^{(2)}_{2c}+2c^{(2)}_{2d}=c^{(2)}_{2c}+4c^{(2)}_{2e}\!,
\end{equation}
\end{subequations}
which replace Eq.\ (\ref{c^(2)-eq3}).
Despite the increase in the number of equations, 
the solution is still given by Eq.\ (\ref{c^(2)}), as seen easily by substituting it into
Eq.\ (\ref{c^(2)-eq-relation2}).
Thus, functional (\ref{Phi^(2)}) with Eq.\  (\ref{c^(2)}) satisfies the Hugenholtz-Pines relation (\ref{relation2})
besides exact relation (\ref{relation1}) for the interaction energy.

 \begin{figure}[t]
        \begin{center}
                \includegraphics[width=0.9\linewidth]{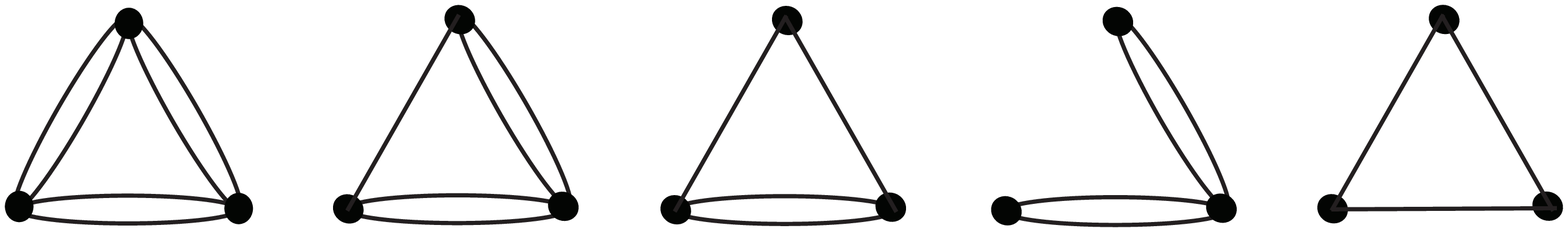}
        \end{center}
        \caption{Distinct diagrams for $\Phi^{(3)}$ without arrows.}
        \label{fig:Phi^(3)}
\end{figure}

\subsection{Constructing $\Phi^{(3)}$}

\begin{figure}[b]
\begin{center}
\includegraphics[width=0.4\linewidth]{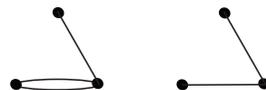}
\end{center}
\caption{Two additional diagrams to calculate Eq.\ (\ref{relation1}) or Eq.\ (\ref{relation2}) for $n=3$.}
\label{fig:Phi^(3)-A}
\end{figure}

The same procedure has been used to obtain the expression of the third-order contribution $\Phi^{(3)}$.
The relevant diagrams are given in Fig.\ \ref{fig:Phi^(3)} without arrows, and Fig.\ \ref{fig:Phi^(3)-A}
shows additional diagrams necessary for the evaluation of Eq.\ (\ref{relation1}) or Eq.\ (\ref{relation2}).
The two relations have been checked to yield a unique result, which is summarized in Appendix B.

\section{Expressions of Entropy and Superfluid Density\label{sec:Eq-Prop}}

The analysis of the preceding section has clarified that we can generally express the thermodynamic potential of
interacting condensed bosons as Eq.\ (\ref{Omega}), and $\Phi$ can be constructed order by order uniquely
so as to satisfy both of exact relations (\ref{relation1}) and (\ref{relation2}).
The fact implies that Eq.\ (\ref{Omega}) becomes exact when terms up to $n=\infty$ are retained in $\Phi$. 
Using Eq.\ (\ref{Omega}), we now derive formally exact expressions of entropy and superfluid density
in terms of Green's function (\ref{hatG}), which are also valid within the gapless $\Phi$-derivable approximation.
We set $\Psi(1)\rightarrow \Psi({\bf r}_{1})$ and $\bar{\Psi}(1)\rightarrow \Psi^{*}({\bf r}_{1})$ below as they have no explicit temperature dependence.

\subsection{Entropy}

Let us expand every quantity in Eq.\ (\ref{Omega}) as
\begin{equation}
\hat{G}(1,2)=T\sum_{n=-\infty}^{\infty}
\hat{G}({\bf r}_{1},{\bf r}_{2};z_{n})e^{-z_{n}(\tau_{1}-\tau_{2})} ,
\label{hatG-exp1}
\end{equation}
for example, with $z_{n}\equiv 2n\pi i T$.
We next transform the summation over $z_{n}$ into an integration on the complex $z$ plane
by using the Bose distribution function: \cite{LW60,AGD63}
\begin{equation}
\hat{f}(z)=\left[
\begin{array}{cc}f(z) & 0
\\
0 & -f(-z)
\end{array}
\right], \hspace{5mm} f(z)\equiv \frac{1}{e^{z/T}-1} .
\label{f(z)}
\end{equation}
The signs of $\pm f(\pm z)$ correspond to the subscripts in Eq.\ (\ref{Tr-def}); with this choice we can deform the original integration contour encircling the imaginary axis down around the real axis. \cite{LW60,AGD63} Note $-f(-z)=1+f(z)$.
Equation (\ref{Omega}) is thereby transformed into
\begin{eqnarray}
&&\hspace{-9mm}
\Omega =\int d^{3}r_{1} \Psi^{*}({\bf r}_{1})K_{1}\Psi({\bf r}_{1})
\nonumber \\
&&\hspace{-2mm}
- {\cal P} \int_{-\infty}^{\infty}\frac{d\varepsilon}{2\pi }{\rm Tr}\hat{f}(\varepsilon) \bigl\{{\rm Im}
\ln [\hat{K}+\hat{\Sigma}(\varepsilon_{-})-\varepsilon_{-}\hat{\sigma}_{0}]
\nonumber \\
&&\hspace{-2mm}
+{\rm Im}\hat{\Sigma}(\varepsilon_{-}) {\rm Re}\hat{G}(\varepsilon_{-})
+{\rm Re}\hat{\Sigma}(\varepsilon_{-}) {\rm Im}\hat{G}(\varepsilon_{-})\bigr\}  
+\Phi .
\label{Omega2}
\end{eqnarray}
Here ${\cal P}$ denotes Cauchy principal value to remove the pole $\varepsilon=0$ of $f(\varepsilon)$
which belongs to $\{z_{n}\}_{n}$, 
Tr is defined as Eq.\ (\ref{Tr-def}) without $\tau$ integrals, 
$\hat{K}\equiv \hat{K}({\bf r}_{1},{\bf r}_{2})$ and $\hat{G}(\varepsilon_{-})
=\hat{G}({\bf r}_{1},{\bf r}_{2};\varepsilon_{-})$ with $\hat{K}({\bf r}_{1},{\bf r}_{2})\equiv \hat{\sigma}_{3}K_{1}\delta ({\bf r}_{1}-{\bf r}_{2})$ and 
$\varepsilon_{-}\equiv \varepsilon+i0_{-}$, and ${\rm Re}\hat{G}(\varepsilon_{-})$ and ${\rm Im}\hat{G}(\varepsilon_{-})$ are defined by 
\begin{subequations}
\begin{eqnarray}
&&{\rm Re}\hat{G}(\varepsilon_{-})\equiv \frac{\hat{G}(\varepsilon_{-})+\hat{G}(\varepsilon_{+})}{2},
\\
&&{\rm Im}\hat{G}(\varepsilon_{-})\equiv \frac{\hat{G}(\varepsilon_{-})-\hat{G}(\varepsilon_{+})}{2i} .
\end{eqnarray}
\end{subequations}
We can also express $\Phi$ in terms of $f(\varepsilon)$
by using the Lehmann representation of $\hat{G}$. \cite{AGD63}
Just like case of the normal system, \cite{Kita99,Kita06Entropy}
inspection of the self-consistent perturbation series obtained in Sec.\ \ref{sec:Phi}
indicates that $\Phi^{(n)}$ and $\hat{\Sigma}^{(n)}$ 
($n=1,2,3$) satisfy 
\begin{equation}
\frac{\delta \Phi}{\delta f(\varepsilon)}
=\frac{1}{2\pi}{\rm Tr}\,{\rm Re}\hat{\Sigma}(\varepsilon_{-}){\rm Im}\hat{G}(\varepsilon_{-}),
\label{dPhidf}
\end{equation}
order by order, which is apparently connected with Eq.\ (\ref{Sigma-Phi}). 
We will proceed by assuming that Eq.\ (\ref{dPhidf}) holds generally.

We now calculate entropy $S=-(\partial \Omega/\partial T)$. 
Noting Eq.\ (\ref{stationarity}), the differentiation needs to be carried out only with respect to the explicit $T$ dependence in $f$. \cite{Kita99}
It then follows from Eq.\ (\ref{dPhidf}) that $-\partial \Phi/\partial T$ exactly cancels the contribution of the third term in the curly bracket of Eq.\ (\ref{Omega2}). 
We also use the relation $\partial f/\partial T=-\partial \sigma/\partial \varepsilon$ with
\begin{equation}
\sigma(\varepsilon)\equiv -f(\varepsilon)\ln |f(\varepsilon)|-f(-\varepsilon)\ln |f(-\varepsilon)| ,
\end{equation}
to perform a partial integration over $\varepsilon$.
We thereby obtain
\begin{eqnarray}
&&\hspace{-6mm}
S ={\cal P}  \int_{-\infty}^{\infty}\frac{d\varepsilon}{2\pi }\sigma(\varepsilon)
{\rm Tr} \biggl\{\biggl[\hat{\sigma}_{0}-
\frac{\partial {\rm Re}\hat{\Sigma}(\varepsilon_{-})}{\partial \varepsilon}\biggr]{\rm Im}\hat{G}(\varepsilon_{-})
\nonumber \\
&&\hspace{0.5mm}
+{\rm Im}\hat{\Sigma}(\varepsilon_{-})\frac{\partial {\rm Re}\hat{G}(\varepsilon_{-})}{\partial \varepsilon}\biggr] .
\label{entropy}
\end{eqnarray}
The expression is a direct extension of the normal-state entropy \cite{Kita99,Kita06Entropy} to
the system of interacting bosons with broken U(1) symmetry. 

Adopting the mean-field approximation without the
$\varepsilon$ dependence in the self-energy $\hat{\Sigma}$,
Eq.\ (\ref{entropy}) reduces to a well-known expression.
To see this, let us diagonalize the operator $\hat{\cal H}\equiv \hat{K}+\hat{\Sigma}$ in 
$\hat{G}^{-1}$ with the Bogoliubov-de Gennes equation: \cite{Kita06}
\begin{equation}
\int \hat{\cal H}({\bf r}_{1},{\bf r}_{2})\hat{u}_{\nu}({\bf r}_{2})\,d^{3}r_{2}
=\hat{u}_{\nu}({\bf r}_{1})\hat{\sigma}_{3}E_{\nu} ,
\label{BdG}
\end{equation}
where $E_{\nu}>0$, and the eigenfunction $\hat{u}_{\nu}({\bf r})$ can be put into the expression:
\begin{equation}
\hat{u}_{\nu}({\bf r})\equiv \left[\begin{array}{cc}
u_{\nu}({\bf r}) & v_{\nu}({\bf r}) 
\\
-v_{\nu}^{*}({\bf r}) & -u_{\nu}^{*}({\bf r})
\end{array}
\right] ,
\label{hatu}
\end{equation}
with $\int d^{3}r\hat{\sigma}_{3}\hat{u}_{\nu'}^{\dagger}({\bf r})\hat{\sigma}_{3}\hat{u}_{\nu}({\bf r})=\hat{\sigma}_{0}\delta_{\nu'\nu}$.
Note that Eq.\ (\ref{GP}) for the condensate wave function is obtained from Eq.\ (\ref{BdG}) as the limit of 
$u_{\nu}$, $v_{\nu}\rightarrow \Psi$, and $E_{\nu}\rightarrow 0$; hence there is no energy gap in the excitation energy
in accordance with Goldstone's theorem.
Green's function is then transformed into
$\hat{G}({\bf r}_{1},{\bf r}_{2};\varepsilon_{-})=
\sum_{\nu}\hat{u}_{\nu}({\bf r}_{1})(\varepsilon_{-}\hat{\sigma}_{0}-E_{\nu}\hat{\sigma}_{3})^{-1}
\hat{\sigma}_{3}\hat{u}_{\nu}^{\rm \dagger}({\bf r}_{2})\hat{\sigma}_{3}$.
Substituting it into Eq.\ (\ref{entropy}) and using the orthonormality of $\hat{u}_{\nu}$, 
we arrive at
$S =\sum_{\nu}[-f_{\nu}\ln f_{\nu}+(1+f_{\nu})\ln (1+f_{\nu})]$ with $f_{\nu}=f(E_{\nu})$.

The above consideration with the mean-field approximation has exemplified that
the structure of $\hat{G}(\varepsilon_{-})$ near $\varepsilon=0$ is directly relevant to
the entropy of BEC at low temperatures. 
It also tells us that the Bogoliubov mode will be connected continuously to the quasiparticle mode which dominates
the low-temperature thermal properties of superfluid $^{4}$He. \cite{Bennemann76}
Further investigations seem required about the general properties of the self-energy near $\varepsilon=0$
to elucidate this connection.

\subsection{Superfluid density}

We next derive an expression of the superfluid density.
Consider a homogeneous BEC where the condensate wave function is given by
$\Psi({\bf r}_{1})=\sqrt{n_{0}}\,e^{i{\bf q}\cdot{\bf r}_{1}}$ with $n_{0}$ as the condensate density.
In this case, the off-diagonal self-energy $\Delta(1,2)$ acquires
the spatial dependence $e^{i{\bf q}\cdot({\bf r}_{1}+{\bf r}_{2})}$ in terms of the 
center-of-mass coordinate $({\bf r}_{1}+{\bf r}_{2})/2$. 
This may be realized by looking at the condensate contribution to $\Delta(1,2)$
which is given by $V({\bf r}_{1}-{\bf r}_{2})\delta(\tau_{1}-\tau_{2})\Psi({\bf r}_{1})\Psi({\bf r}_{2})$. 
It hence follows that Green's function can be expanded in terms of the basis function:
\begin{equation}
\hat{\varphi}_{\bf p}({\bf r}_{1})\equiv \frac{e^{i{\bf p}\cdot{\bf r}_{1}}}{\sqrt{\cal V}}
\left[\begin{array}{cc}
\vspace{1mm}
e^{i{\bf q}\cdot{\bf r}_{1}} & 0
\\
0 & e^{-i{\bf q}\cdot{\bf r}_{1}}
\end{array}\right] ,
\end{equation}
with ${\cal V}$ as the volume of the system, as 
\begin{equation}
\hat{G}(1,2)=T\sum_{n{\bf p}}\hat{\varphi}_{\bf p}({\bf r}_{1})
\hat{G}_{\bf p}(z_{n})\hat{\varphi}_{\bf p}^{*}({\bf r}_{2})e^{-z_{n}(\tau_{1}-\tau_{2})} ,
\label{hatG-exp2}
\end{equation}
with $z_{n}=2n\pi iT$.
The momentum density $\langle {\bf p}\rangle$ is calculated by operating $-i{\bf\nabla}_{1}$ to $\langle T_{\tau}\psi(1)\psi^{\dagger}(2)\rangle =\Psi({\bf r}_1)\Psi^{*}({\bf r}_{2})-G(1,2)$ and setting
$2=1_{+}$ subsequently. Noting $\bar{G}(1,2)=G(2,1)$ in Eq.\ (\ref{hatG}), the result can also be expressed in terms of the Nambu matrix $\hat{G}_{\bf p}(z_{n})$
in Eq.\ (\ref{hatG-exp2}) as
\begin{eqnarray}
&&\hspace{-4mm}
\langle {\bf p}\rangle =n_{0}{\bf q}-\frac{T}{2{\cal V}}\sum_{n{\bf p}}{\rm Tr}
\!\left[\begin{array}{cc}
\vspace{1mm}
{\bf p}+\!{\bf q} & 0
\\
0 & -{\bf p}+\!{\bf q}
\end{array}\right]\!
\hat{\sigma}_{3}
\hat{G}_{\bf p}(z_{n})\hat{1}(z_{n})
\nonumber \\
&&\hspace{2mm}
=n{\bf q}
-\frac{T}{2{\cal V}}\sum_{n{\bf p}}{\bf p}\,{\rm Tr}\,
\hat{G}_{\bf p}(z_{n})\hat{1}(z_{n}) .
\label{<p>}
\end{eqnarray}
Here $\hat{1}(z_{n})$ is defined by
\begin{equation}
\hat{1}(z_{n})\equiv \left[\begin{array}{cc}
\vspace{1mm}
e^{z_{n}0_{+}} & 0
\\
0 & e^{z_{n}0_{-}}
\end{array}\right] ,
\end{equation}
and we have used $n_{0}-G(1,1_{+})=n$ with $n$ denoting the particle density.
We further express the term with $\hat{G}_{\bf p}(z_{n})$ in Eq.\ (\ref{<p>}) as
\begin{eqnarray*}
&&\hspace{-5mm}
\sum_{\bf p}{\bf p}\,{\rm Tr}\,
\hat{G}_{\bf p}(z_{n})\hat{1}(z_{n})=\sum_{\bf p}{\bf p}\,{\rm Tr}\,
\hat{1}(z_{n})\frac{\partial}{\partial z_{n}}\ln[-\hat{G}_{\bf p}^{-1}(z_{n})]
\nonumber \\
&&\hspace{30mm}+\sum_{\bf p}{\bf p}\,{\rm Tr}\,
\hat{1}(z_{n})\frac{\partial \hat{\Sigma}_{\bf p}(z_{n})}{\partial z_{n}}\hat{G}_{\bf p}(z_{n}) .
\end{eqnarray*}
It then turns out that the second term on the right-hand side vanishes, as can be
seen by repeating the argument from Eq.\ (3.18) to Eq.\ (3.23) in Ref.\ \onlinecite{Kita96}
based on the momentum conservation of $\Phi$ at each interaction for homogeneous systems. 
On the other hand, the first term can be transformed with the Bose distribution function $f$ into an integration
just below and above the real axis on the complex $z$ plane.
Carrying out a partial integration subsequently, 
we obtain an expression of the superfluid density tensor $\rho^{({\rm s})}_{ij}\equiv
m(\partial \langle p_{i}\rangle/\partial q_{j})_{{\bf q}={\bf 0} }$ as
\begin{eqnarray}
&&\hspace{-5mm}
\rho^{({\rm s})}_{ij}=mn\delta_{ij}-\frac{m}{{\cal V}}\sum_{{\bf p}}{\cal P} \int_{-\infty}^{\infty}\frac{d\varepsilon}{2\pi}
\frac{\partial f(\varepsilon)}{\partial\varepsilon}
\nonumber \\
&&\hspace{4.7mm}\times
p_{i}\frac{\partial}{\partial q_{j}}
{\rm Tr}\,{\rm Im}
\ln[-\hat{G}_{\bf p}^{-1}(\varepsilon_{-})]\biggr|_{{\bf q}={\bf 0}}.
\label{rho_s}
\end{eqnarray}
This expression manifestly tells us that 
the structure of $\hat{G}_{\bf p}(\varepsilon_{-})$ near $\varepsilon=0$ is directly relevant to
the superfluid density.
Without $\varepsilon$ and ${\bf p}$ dependences in the self-energy, for example, 
the formula reproduces the expression given
by Fetter \cite{Fetter72} for the weakly interacting condensed Bose gas.
It is also worth pointing out that Eq.\ (\ref{rho_s}) is identical in form with that of Fermi superfluids \cite{Kita96}
including the BCS-BEC crossover regime. \cite{FOTG07}

\section{Real-time equations of motion\label{sec:real-time}}

The formulation of Secs.\ \ref{sec:exact_relations} and \ref{sec:Phi} can be extended straightforwardly to describe nonequilibrium dynamical evolutions of BEC. 
Formally, we only need to replace the integration contour $0\leq \tau\leq T^{-1}$ by the
closed time-path contour. \cite{CSHY85,RS86}
We here follow Keldysh \cite{Keldysh64,RS86} to distinguish the forward and backward branches
so that every time integration is limited to $-\infty < t<\infty$.
The present transformation from equilibrium to nonequilibrium 
is a direct extension of the normal-state consideration. \cite{Kita06Entropy}

Let us replace $1\rightarrow 1^{j}\equiv ({\bf r}_{1},t_{1}^{j})$ and $\tau_{1}\rightarrow it_{1}^{j}$ in the Heisenberg operators of 
Eq.\ (\ref{psi(1)-def}), where superscript $j=1,2$ distinguishes the forward $(j=1)$ and backward $(j=2)$ branches.
Using them, we next define Green's function in the Nambu space by
\begin{eqnarray}
&&\hspace{-5mm}
\hat{G}_{ij}(1,2)\equiv - i\left< T_{C}\!
\left[
\begin{array}{c}
\vspace{1mm}
\phi (1^{i})\\ \bar{\phi}(1^{i})
\end{array}
\right] \! [\, \bar{\phi}(2^{j}) \,\, \phi(2^{j})\,] \right> \hat{\sigma}_{3}
\nonumber \\
&&\hspace{9mm}\equiv \left[
\begin{array}{cc}
\vspace{1mm}
G_{ij}(1,2) & F_{ij}(1,2)
\\
-\bar{F}_{ij}(1,2) & -\bar{G}_{ij}(1,2)
\end{array}\right] ,
\label{hatG-r}
\end{eqnarray}
with $T_C$ as the generalized time-ordering operator. \cite{Keldysh64,RS86}
The elements obey
$G_{ij}^{*}(1,2)=-G_{3-j,3-i}(2,1)$, $F_{ij}(1,2)=F_{ji}(2,1)$, $\bar{F}_{ij}(1,2)=-F_{3-j,3-i}^{*}(2,1)$, and $\bar{G}_{ij}(1,2)=G_{ji}(2,1)$.
The self-energy $\hat{\Sigma}_{ij}(1,2)$ is defined similarly as Eq.\ (\ref{hatSigma}) with the additional subscripts ${ij}$.
We next introduce the $4\times 4$ matrices (distinguished with $\check{\hspace{1mm}}$ on top):
\begin{subequations}
\begin{equation}
\check{G}(1,2)\equiv 
\left[\begin{array}{cc} 
\vspace{1mm}
\hat{G}_{11}(1,2) & \hat{G}_{12}(1,2) 
\\
\hat{G}_{21}(1,2) & \hat{G}_{22}(1,2) 
\end{array}\right] , 
\label{checkG}
\end{equation}
\begin{equation}
\check{G}_{0}^{-1}(1,2)\equiv 
\left[\begin{array}{cc}
\vspace{1mm}
\hat{G}_{0}^{-1}(1,2) & \hat{0}
\\
\hat{0} & - \hat{G}_{0}^{-1}(1,2) 
\end{array}
\right] ,
\label{checkG_0^-1}
\end{equation}
\begin{equation}
\check{\sigma}_{0}\equiv 
\left[\begin{array}{cc} 
\vspace{1mm}
\hat{\sigma}_{0} & \hat{0}
\\
\hat{0} & \hat{\sigma}_{0} 
\end{array}\right] ,\hspace{5mm}
\check{\sigma}_{3}\equiv 
\left[\begin{array}{cc} 
\vspace{1mm}
\hat{\sigma}_{0} & \hat{0}
\\
\hat{0} & -\hat{\sigma}_{0} 
\end{array}\right] ,
\end{equation}
\end{subequations}
where $\hat{G}_{0}^{-1}(1,2)$ is
given by Eq.\ (\ref{hatG_0^-1}) with ${\partial}/{\partial\tau_{1}}\rightarrow -i{\partial}/{\partial t_{1}}$. The self-energy matrix $\check{\Sigma}$ is defined in the same way as Eq.\ (\ref{checkG}). 
Now, the Dyson-Beliaev equation reads as
\begin{equation}
\int d3 [\check{G}_{0}^{-1}(1,3)-\check{\sigma}_{3} \check{\Sigma}(1,3)\check{\sigma}_{3}]\check{G}(3,2)
=\delta(1,2)\check{\sigma}_{0},
\label{Dyson-r}
\end{equation}
where two $\check{\sigma}_{3}$'s originate from the path inversion for $j=2$. \cite{Kita06Entropy}
Also, Eq.\ (\ref{GP}) for the condensate wave function is replaced by
\begin{equation}
\int \! d2 \biggl[\hat{G}_{0}^{-1}(1,2)-\sum_{j=1}^{2}\hat{\Sigma}_{1j}(1,2)(-1)^{j-1}\biggr]\!\!
\left[\!\begin{array}{c}
\vspace{1mm}
\Psi(2)
\\
-\bar{\Psi}(2)
\end{array}
\!\right] \!=\!
\left[
\begin{array}{c}
0 \\ 0\end{array}\!\right] \! ,
\label{GP-r}
\end{equation}
where we have incorporated $\Psi(1)\equiv \Psi(1^{1})=\Psi(1^{2})$.
Note $\bar{\Psi}(1)=\Psi^{*}(1)$ here.

Green's function in Eq.\ (\ref{hatG-r}) contains an extra factor $i$ compared with Eq.\ (\ref{hatG}).
Taking this fact into account, the real-time functional $\Phi^{(n)}$ is obtained from the equilibrium one of Sec.\ \ref{sec:Phi} with the following modifications:
(i) add branch indices to Green's function and the vertex [Eq.\ (\ref{Gamma^(0)})] as 
$\hat{G}\rightarrow \hat{G}_{ij}$  and
\begin{eqnarray}
&&\hspace{-5mm}
\Gamma^{(0)}_{ii',jj'}(11',22')
\equiv (-1)^{j-1}\delta_{ij}\delta_{ii'}\delta_{jj'}V({\bf r}_{1}-{\bf r}_{2})\delta(t_{1}-t_{2})
\nonumber \\
&&\hspace{22.5mm}
\times[\delta(1,1')\delta(2,2')+\delta(1,2')\delta(2,1')] ,
\label{Gamma^(0)-r}
\end{eqnarray}
respectively, where the factor $(-1)^{j-1}$ is due to the path inversion for $j=2$;
(ii) include summations over the branch indices; (iii) multiply each term by $i^{m/2}$, 
where $m$ is the number of condensate wave functions in the relevant term;
(iv) multiply the resultant expression by $(-1)^{n-1}i^{n}/T$. 
In steps (iii) and (iv), we have removed the factor $T$
in equilibrium $\Phi$ originating from $\Omega=-T\ln{\rm Tr}\,e^{-H/T}$,
and also considered the path change $\tau_{1}\rightarrow it_{1}^{j}$ to reproduce the 
overall factor $(-i)^{n}$ in the $n$th-order perturbation.

Thus, Eq.\ (\ref{Phi^(1)}) is now replaced by
\begin{eqnarray}
&&\hspace{-4mm}
\Phi^{(1)}
=\frac{i}{4} \sum_{ii'jj'}\int d1 \int d1' \int d2 \int d2'\,\Gamma^{(0)}_{ii',jj'}(11',22')
\nonumber \\
&&\hspace{7mm}
\times \bigl\{ 2G_{ii'}(1,1')G_{jj'}(2,2')\!
+c^{(1)}_{2b}F_{ij}(1,2) \bar{F}_{i'j'}(1',2')
\nonumber \\
&&\hspace{7mm}
+ic^{(1)}_{1a}G_{ii'}(1,1')\Psi_{j}(2)\bar{\Psi}_{j'}(2')
+ic^{(1)}_{1b}\bigl[F_{ij}(1,2)
\nonumber \\
&&\hspace{7mm}
\times \bar{\Psi}_{i'}(1')\bar{\Psi}_{j'}(2')
+\bar{F}_{i'j'}(1',2')
 \Psi_{i}(1)\Psi_{j}(2)\bigr]
\nonumber \\
&&\hspace{7mm}
+i^{2}\bar{\Psi}_{i'}(1')\bar{\Psi}_{j'}(2')\Psi_{j}(2)\Psi_{i}(1) \bigr\} ,
\label{Phi^(1)-r}
\end{eqnarray}
where $ \Psi_{i}(1)\equiv\Psi(1^{i})$, and $c^{(1)}_{\nu}$s are given as Eq.\ (\ref{c^(1)}).
Real-time functionals $\Phi^{(2)}$ and $\Phi^{(3)}$ can be constructed similarly from
Eqs.\ (\ref{Phi^(2)}) and (\ref{Phi^(3)}), respectively, with the coefficients of Eqs.\ (\ref{c^(2)}) and (\ref{c^(3)}).
Using $G_{ij}^{*}(1,2)=-G_{3-j3-i}(2,1)$,
$\bar{F}_{ij}(1,2)=-F_{3-j3-i}^{*}(2,1)$, and 
$\Gamma^{(0)*}_{ii',jj'}(11',22')=-\Gamma^{(0)}_{3-i3-i',3-j3-j'}(11',22')$,
 one can show $\Phi^{(n)*}=\Phi^{(n)}$.

Accordingly, Eq.\ (\ref{Sigma-Phi}) for the self-energies are modified into
\begin{subequations}
\label{Sigma-r}
\begin{equation}
\Sigma_{ij}(1,2)=(-1)^{i+j}\frac{\delta \Phi}{\delta G_{ji}(2,1)},
\end{equation}
\begin{equation}
\Delta_{ij}(1,2) =-2(-1)^{i+j}\frac{\delta \Phi}{\delta\bar{F}_{ji}(2,1)},
\end{equation}
\end{subequations}
where the factor $(-1)^{i+j}$ is due to the two $\check{\sigma}_{3}$'s in Eq.\ (\ref{Dyson-r}).
It follows from $\Phi^{(n)*}=\Phi^{(n)}$ and the symmetries of $\check{G}$
that $\Sigma_{ij}^{*}(1,2)=-\Sigma_{3-j,3-i}(2,1)$, $\Delta_{ij}(1,2)=\Delta_{ji}(2,1)$, $\bar{\Delta}_{ij}(1,2)=-\Delta_{3-j,3-i}^{*}(2,1)$, and $\bar{\Sigma}_{ij}(1,2)=\Sigma_{ji}(2,1)$. The functional also satisfies
\begin{eqnarray}
&&\hspace{-5mm}
i\frac{\delta\Phi}{\delta \bar{\Psi}_{i}(1)}
=\sum_{j}(-1)^{i+j}\int d2 \bigl[\Sigma_{ij}(1,2)\Psi_{j}(2)
\nonumber \\
&&\hspace{11.5mm}-\Delta_{ij}(1,2)\bar{\Psi}_{j}(2)\bigr],
\end{eqnarray}
which corresponds to Eq.\ (\ref{Psi-Phi}).

Equations (\ref{Dyson-r}), (\ref{GP-r}), and (\ref{Sigma-r}) form self-consistent equations for nonequilibrium time evolutions
of BEC satisfying conservation laws and Goldstone's theorem simultaneously.
Approximating $\Phi$ by $\Phi^{(1)}+\Phi^{(2)}$ yields a non-vanishing collision integral, for example. 
Using it, we can describe thermalization of weakly interacting BEC microscopically, i.e., a topic which seems not to have been clarified sufficiently. See, e.g., a recent book by Griffin, Nikuni, 
and Zaremba \cite{GNZ09} for the present status on this issue.

\section{Summary\label{sec:summary}}

We have developed a self-consistent perturbation expansion for BEC with broken U(1) symmetry
so as to obey Goldstone's theorem and dynamical conservation laws simultaneously.
First, the Luttinger-Ward thermodynamic functional for the normal state \cite{LW60} has been extended to 
a system of interacting condensed bosons as Eq.\ (\ref{Omega}).
Next, we have presented a procedure to construct $\Phi$ in the functional order by order with
exact relations (\ref{relation1}) and (\ref{relation2}) in Sec.\ \ref{subsec:Proc-Phi}.
It has been shown subsequently up to the third order of the self-consistent perturbation expansion  that
both of the relations yield a unique identical result at each order as
Eq.\ (\ref{Phi^(1)}) with Eq.\ (\ref{c^(1)}), Eq.\ (\ref{Phi^(2)}) with Eq.\ (\ref{c^(2)}), 
and Eq.\ (\ref{Phi^(3)}) with Eq.\ (\ref{c^(3)}). 
This fact implies that the expansion converges to the exact thermodynamic potential when infinite terms 
are retained in $\Phi$.
Using Eq.\ (\ref{Omega}), 
we have also derived useful expressions for the entropy and superfluid density in terms of Green's function  as Eqs.\ (\ref{entropy})
and (\ref{rho_s}), respectively.
Finally, we have derived a set of real-time dynamical equations for BEC as Eqs.\ (\ref{Dyson-r}), (\ref{GP-r}),
and (\ref{Sigma-r}).

An expansion scheme like the present one may have been anticipated
since the work of de Dominicis and Martin  \cite{dDM64} in 1964 to prove the existence of the functional satisfying Eq.\ (\ref{stationarity}).
However, no explicit expression for the functional seems to have been known to date. 
As already noted in Introduction, one of the advantages of the present expansion scheme 
over the simple perturbation expansion lies in its ability to describe nonequilibrium phenomena.
Another point to be mentioned is that it incorporates 
effects of the anomalous Green's function 
$F(1,2)=\langle T_{\tau}\phi(1)\phi(2)\rangle$
more efficiently than the simple perturbation expansion. \cite{AGD63}
This fact may be realized by noting that $F(1,2)$ becomes finite with at least a single interaction line 
in the latter scheme. Thus, $n$th-order terms with $F$ or $\bar{F}$ in the present expansion
contain effects which show up only after the $(n\!+\!1)$th order in the simple perturbation expansion.

We are planning to apply the present formalism to a wide range of systems/phenomena in BEC
to elucidate their properties microscopically. It also remains to be performed to clarify
two-particle correlations within the present formalism.

\appendix

\section{Derivation of Eqs.\ (\ref{Dyson}), (\ref{GP}), and (\ref{<H_int>})}

Following the procedure sketched by Hohenberg and Martin, \cite{HM65} 
we here derive the Dyson-Beliaev Eq.\  (\ref{Dyson}) and 
the Hugenholtz-Pines relation (\ref{GP}) for general inhomogeneous systems
so as to be compatible with our definition [Eq.\ (\ref{hatG})] of Green's function.
We also prove expression (\ref{<H_int>})
for the interaction energy.

\subsection{Time evolution operator\label{subsec:TEO}}

Let us introduce the external perturbation: \cite{dDM64,HM65,CJT74}
\begin{equation}
H_{\rm ext}(\tau_{1})\equiv \int d^{3}r_{1} \left[\psi^{\dagger}({\bf r}_{1})\eta_{\rm ext}(1)
+\psi({\bf r}_{1})\eta_{\rm ext}^{*}(1)\right] ,
\label{source}
\end{equation}
where $\eta_{\rm ext}(1)$ is periodic in $\tau_{1}$ with the period
$T^{-1}$.
The total Hamiltonian in this Appendix is given as a sum of Eqs.\ (\ref{Hamil}) and (\ref{source}) by
\begin{equation}
{\cal H}(\tau_{1})\equiv H+H_{\rm ext}(\tau_{1}) .
\label{calH}
\end{equation}
The extra term $H_{\rm ext}$ serves as a convenient tool to derive various formal relations.
The limit $\eta_{\rm ext}\rightarrow 0$ will be taken once all the necessary formulas are obtained.

We next define a time evolution operator in terms of ${\cal H}$ by
\begin{eqnarray}
&&\hspace{-1mm}
{\cal U}(\tau,\tau_{0})
\nonumber \\
&&\hspace{-5mm}
\equiv 1+\sum_{n=1}^{\infty}(-1)^{n}\int_{\tau_{0}}^{\tau}d \tau_{n}\cdots
\int_{\tau_{0}}^{\tau_{2}}d \tau_{1}{\cal H}(\tau_{n})\cdots{\cal H}(\tau_{1})
\nonumber \\
&&\hspace{-5mm}=\left\{
\begin{array}{ll}\vspace{2mm}\displaystyle
T_{\tau}\exp\left[-\int_{\tau_{0}}^{\tau}d \tau_{1}{\cal H}(\tau_{1}) \right] & :\tau\geq {\tau_{0}}
\\
\displaystyle
T_{\tau}^{a}\exp\left[-\int_{\tau_{0}}^{\tau}d \tau_{1}{\cal H}(\tau_{1}) \right] & :\tau<  {\tau_{0}}
\end{array}\right. ,
\label{calU}
\end{eqnarray}
where $T_{\tau}^{a}$ is the anti-time-ordering operator.
Note ${\cal U}(\tau,\tau_{0})\rightarrow e^{-(\tau-\tau_{0})H}$ as $\eta_{\rm ext}\rightarrow 0$.
This operator ${\cal U}(\tau,\tau_{0})$ obeys
\begin{subequations}
\label{calU-eq-motion}
\begin{equation}
\frac{d\, {\cal U}(\tau,\tau_{0})}{d \tau}=-{\cal H}(\tau){\cal U}(\tau,\tau_{0}) ,
\end{equation}
\begin{equation}
\frac{d\, {\cal U}(\tau,\tau_{0})}{d \tau_{0}}={\cal U}(\tau,\tau_{0}){\cal H}(\tau_{0}).
\label{U-eq}
\end{equation}
\end{subequations}
It also satisfies ${\cal U}(\tau_{0},\tau_{0})=1$ and
\begin{equation}
{\cal U}(\tau,\tau_{1}){\cal U}(\tau_{1},\tau_{0})={\cal U}(\tau,\tau_{0}) .
\label{U-Identity}
\end{equation}
Equation (\ref{U-Identity}) is proved as follows.
We see easily that $\tilde{\cal U}(\tau,\tau_{0}) \equiv {\cal U}(\tau,\tau_{1}){\cal U}(\tau_{1},\tau_{0})$ obeys the same first-order differential equation with respect to $\tau$ as ${\cal U}(\tau,\tau_{0})$.
We also notice that the initial values at $\tau=\tau_{1}$ are the same between the two operators, i.e.,
$\tilde{\cal U}(\tau_{1},\tau_{0})={\cal U}(\tau_{1},\tau_{0})$.
We hence conclude Eq.\ (\ref{U-Identity}). 
Note especially that ${\cal U}^{-1}(\tau,\tau_{0})={\cal U}(\tau_{0},\tau)$, as can be seen easily by setting 
$\tau_{0}=\tau$ in Eq.\ (\ref{U-Identity}).

\subsection{Equations of motion\label{subsec:Eq-motion}}

We now introduce the Heisenberg operators:
\begin{equation}
\left.
\begin{array}{l}
\vspace{2mm}
\psi(1)\equiv {\cal U}^{-1}(\tau_{1})\psi({\bf r}_{1}){\cal U}(\tau_{1}),
\\
\bar{\psi}(1)\equiv {\cal U}^{-1}(\tau_{1})\psi^{\dagger}({\bf r}_{1}){\cal U}(\tau_{1}),
\end{array}\right. 
\label{IR}
\end{equation}
where ${\cal U}(\tau_{1})\equiv {\cal U}(\tau_{1},0)$ and ${\cal U}^{-1}(\tau_{1})\equiv {\cal U}(0,\tau_{1})$
with $0\leq \tau_{1}\leq T^{-1}$.
Differentiating them with respect to $\tau_{1}$ and using Eq.\ (\ref{calU-eq-motion}), we obtain
\begin{subequations}
\label{psi-eq-motion-t}
\begin{eqnarray}
&&\hspace{-5mm}
\biggl(-\frac{\partial}{\partial \tau_{1}}-K_{1}\biggr)\psi(1)
\nonumber \\
&&\hspace{-8mm}
=\eta_{\rm ext}(1)+\int  d 1'\,
\bar{V}(1,1')\bar{\psi}(1')\psi(1')\psi(1),
\label{psi-eq-motion}
\end{eqnarray}
\begin{eqnarray}
&&\hspace{-5mm}
\biggl(\frac{\partial}{\partial \tau_{1}}-K_{1}\biggr)\bar{\psi}(1)
\nonumber \\
&&\hspace{-8mm}
=\eta_{\rm ext}^{*}(1)+ \int  d 1'\, \bar{V}(1,1')
\bar{\psi}(1)\bar{\psi}(1')\psi(1'),
\label{psi^d-eq-motion}
\end{eqnarray}
\end{subequations}
with $\bar{V}(1,1')\equiv \delta(\tau_{1}-\tau_{1}')V({\bf r}_{1}-{\bf r}_{1}')$.

Let us define the expectation value of an arbitrary operator 
${\cal O}(1)\equiv {\cal U}^{-1}(\tau_{1}){\cal O}({\bf r}_{1}){\cal U}(\tau_{1})$ by
\begin{equation}
\langle {\cal O}(1) \rangle\equiv \frac{ {\rm Tr}\,T_{\tau}\,{\cal U}(\beta) {\cal O}(1)}{{\rm Tr}\,{\cal U}(\beta)}
\label{<calO>}
\end{equation}
with $\beta\equiv T^{-1}$,
which for $\eta_{\rm ext}\rightarrow 0$ reduces to the grand-canonical average with respect to $H$.
We then realize from Eq.\ (\ref{psi-eq-motion-t}) that the quantities
\begin{equation}
\Psi(1)\equiv \langle\psi(1)\rangle,\hspace{5mm}\bar{\Psi}(1)\equiv \langle\bar{\psi}(1)\rangle,
\label{<Psi>}
\end{equation}
obey the equations of motion:
\begin{subequations}
\label{Psi-eq-motion-t}
\begin{equation}
\biggl(-\frac{\partial}{\partial \tau_{1}}-K_{1}\biggr)\Psi(1)=\eta_{\rm ext}(1)+\eta(1),
\label{Psi-eq-motion}
\end{equation}
\begin{equation}
\biggl(\frac{\partial}{\partial \tau_{1}}-K_{1}\biggr)\bar{\Psi}(1)
=\eta_{\rm ext}^{*}(1)+\bar{\eta}(1),
\label{Psi^d-eq-motion}
\end{equation}
\end{subequations}
with
\begin{subequations}
\label{eta-def-t}
\begin{equation}
\eta(1)\equiv \int  d 1'\,
\bar{V}(1,1')\langle\bar{\psi}(1')\psi(1')\psi(1)\rangle,
\label{eta-def}
\end{equation}
\begin{equation}
\bar{\eta}(1)\equiv \int  d 1'\, \bar{V}(1,1')
\langle\bar{\psi}(1)\bar{\psi}(1')\psi(1')\rangle .
\label{eta^d-def}
\end{equation}
\end{subequations}

\subsection{Dyson-Beliaev equation\label{subsec:DB-proof}}

To derive Eq.\ (\ref{Dyson}), we first differentiate $\Psi(1)$ in Eq.\ (\ref{<Psi>}) with respect to $\eta_{\rm ext}(2)$.  
Using definition (\ref{<calO>}), one can easily show 
$$
\frac{\delta \Psi(1)}{\delta \eta_{\rm ext}(2)}=-\langle T_{\tau}\psi(1)\bar{\psi}(2) \rangle
+\Psi(1)\bar{\Psi}(2)
=G(1,2),
$$
where $G(1,2)$ is defined by Eq.\ (\ref{hatG}) with definition (\ref{<calO>}) for the expectation value.
Similar calculations lead to
\begin{subequations}
\label{Psi-diff}
\begin{equation}
\frac{\delta \Psi(1)}{\delta \eta_{\rm ext}(2)}=G(1,2),\hspace{5mm}
\frac{\delta \Psi(1)}{\delta \eta_{\rm ext}^{*}(2)}= -F(1,2),
\end{equation}
\begin{equation}
\frac{\delta \bar{\Psi}(1)}{\delta \eta_{\rm ext}(2)}= -\bar{F}(1,2),
\hspace{5mm}
\frac{\delta \bar{\Psi}(1)}{\delta \eta_{\rm ext}^{*}(2)}= \bar{G}(1,2).
\end{equation}
\end{subequations}
We next introduce the self-energies by
\begin{subequations}
\begin{eqnarray}
&&\hspace{-10mm}
\Sigma(1,2)\equiv\frac{\delta  \eta(1)}{\delta \Psi (2)},
\hspace{5mm}
\Delta(1,2)\equiv \frac{\delta  \eta(1)}{\delta \bar{\Psi} (2)},
\end{eqnarray}
\begin{eqnarray}
&&\hspace{-10mm}
\bar{\Delta}(1,2)\equiv \frac{\delta  \bar{\eta}(1)}{\delta \Psi (2)},
\hspace{5mm}
\bar{\Sigma}(1,2)\equiv \frac{\delta  \bar{\eta}(1)}{\delta \bar{\Psi} (2)}.
\end{eqnarray}
\end{subequations}
It then follows that ${\delta \eta(1)}/{\delta\eta_{\rm ext}(2)}$, etc.,  can be expressed as
\begin{subequations}
\label{eta-diff}
\begin{eqnarray}
&&\hspace{-5mm}
\frac{\delta \eta(1)}{\delta\eta_{\rm ext}(2)}=\int d 3
\left[\frac{\delta \eta(1)}{\delta\Psi(3)}\frac{\delta \Psi(3)}{\delta\eta_{\rm ext}(2)}
+\frac{\delta \eta(1)}{\delta\bar{\Psi}(3)}\frac{\delta \bar{\Psi}(3)}{\delta\eta_{\rm ext}(2)}\right]
\nonumber \\
&&\hspace{8.7mm}
=\int d 3\left[\Sigma(1,3)G(3,2)-\Delta(1,3)\bar{F}(3,2)\right],
\nonumber \\
\label{eta-diff1}
\end{eqnarray}
\begin{equation}
\frac{\delta \eta(1)}{\delta\eta_{\rm ext}^{*}(2)}
=\int d 3\left[-\Sigma(1,3)F(3,2)+\Delta(1,3)\bar{G}(3,2)\right],
\end{equation}
\begin{equation}
\frac{\delta \bar{\eta}(1)}{\delta\eta_{\rm ext}(2)}
=\int d 3\left[\bar{\Delta}(1,3)G(3,2)-\bar{\Sigma}(1,3)\bar{F}(3,2)\right] ,
\end{equation}
\begin{equation}
\frac{\delta \bar{\eta}(1)}{\delta\eta_{\rm ext}^{*}(2)}
=\int d 3\left[-\bar{\Delta}(1,3)F(3,2)+\bar{\Sigma}(1,3)\bar{G}(3,2)\right] .
\end{equation}
\end{subequations}
With these preliminaries, we now differentiate Eq.\ (\ref{Psi-eq-motion-t}) with respect to $\eta_{\rm ext}(2)$
or  $\eta^{*}_{\rm ext}(2)$ and set $\eta_{\rm ext}=0$ subsequently.
Using Eqs.\ (\ref{Psi-diff}) and (\ref{eta-diff}), one may see easily that 
the resultant four equations of motion can be written compactly as Eq.\ (\ref{Dyson}).

\subsection{Hugenholtz-Pines relation\label{subsec:HP-proof}}

Equation (\ref{GP}) can be regarded as Goldstone's theorem \cite{PS95,Goldstone61} for the broken U(1)
symmetry. To derive it, we consider the gauge transformation:
\begin{equation}
\eta_{\rm ext}(1)\rightarrow e^{i\chi}\,\eta_{\rm ext}(1),\hspace{5mm}
\psi({\bf r}_{1})\rightarrow e^{i\chi}\psi({\bf r}_{1}) ,
\label{GT}
\end{equation}
where $\chi$ is constant.
This brings first-order changes in various quantities as
$\delta \eta_{\rm ext}(1)=i\chi \eta_{\rm ext}(1)$,
$\delta \eta_{\rm ext}^{*}(1)=-i\chi \eta_{\rm ext}^{*}(1)$,
$\delta \Psi(1)=i\chi \Psi(1)$, and
$\delta \bar{\Psi}(1)=-i\chi \bar{\Psi}^{*}(1)$.
Collecting terms of first order in Eq.\ (\ref{Psi-eq-motion-t}),
we obtain
\begin{subequations}
\label{Psi-eq-motion-1st-t}
\begin{eqnarray}
&&\hspace{-5mm}
0=\biggl(-\frac{\partial}{\partial \tau_{1}}-K_{1}\biggr)\delta\Psi(1)-\delta \eta_{\rm ext}(1)-
\delta \eta(1)
\nonumber \\
&&\hspace{-2.5mm}
=i\chi \biggl\{\biggl(-\frac{\partial}{\partial \tau_{1}}-K_{1}\biggr)\Psi(1)-\eta_{\rm ext}(1)
\nonumber \\
&&\hspace{1mm}
-\int d 2
\left[\Sigma(1,2)\Psi(2)-\Delta(1,2)\bar{\Psi}(2)\right]\biggr\} ,
\label{Psi-eq-motion-1st}
\end{eqnarray}
\begin{eqnarray}
&&\hspace{-5mm}
0=i\chi \biggl\{\biggl(-\frac{\partial}{\partial \tau_{1}}+K_{1}\biggr)\bar{\Psi}(1)+\eta_{\rm ext}^{*}(1)
\nonumber \\
&&\hspace{1mm}
-\int d 2
\left[\bar{\Delta}(1,2)\Psi(2)-\bar{\Sigma}(1,2)\bar{\Psi}(2)\right]\biggr\} .
\label{barPsi-eq-motion-1st}
\end{eqnarray}
\end{subequations}
respectively, where we have performed the same transformation for $\delta \eta(1)$ as Eq.\ (\ref{eta-diff1}).
Noting $\chi$ is arbitrary
and comparing Eqs.\ (\ref{Psi-eq-motion-t}) and (\ref{Psi-eq-motion-1st-t}), we obtain
\begin{subequations}
\label{eta-Sigma}
\begin{equation}
\eta(1)=\int d 2
\left[\Sigma(1,2)\Psi(2)-\Delta(1,2)\bar{\Psi}(2)\right] ,
\end{equation}
\begin{equation}
\bar{\eta}(1)=\int d 2
\left[-\bar{\Delta}(1,2)\Psi(2)+\bar{\Sigma}(1,2)\bar{\Psi}(2)\right] .
\end{equation}
\end{subequations}
We finally substitute Eq.\ (\ref{eta-Sigma}) into
Eq.\ (\ref{Psi-eq-motion-t}) and set $\eta_{\rm ext}=0$.
We thereby arrive at Eq.\ (\ref{GP}).

\subsection{Interaction energy\label{subsec:IntE-proof}}

To derive Eq.\ (\ref{<H_int>}), we multiply Eqs.\ (\ref{psi-eq-motion}) and (\ref{psi^d-eq-motion}) 
with $\eta_{\rm ext}=0$ by $\bar{\phi}(1')$ and $\phi(1')$, 
respectively, operate $T_{\tau}$, and take the thermodynamic average 
with Eq.\ (\ref{psi=Psi+psi-t}) and $\langle\phi\rangle=0$ in mind.
Noting $-\langle T_{\tau}\frac{\partial \phi(1)}{\partial\tau_{1}}\bar{\phi}(1')\rangle=-\frac{\partial}{\partial\tau_{1}}
\langle T_{\tau}\phi(1)\bar{\phi}(1')\rangle+\delta(1,1')$ and its conjugate, we can express
the resultant equations in terms of the diagonal elements of Eq.\ (\ref{hatG}) as
\begin{subequations}
\begin{eqnarray}
&&\hspace{-2mm}
-\biggl(-\frac{\partial}{\partial \tau_{1}}-K_{1}\biggr)G(1,1')+\delta(1,1')
\nonumber \\
&&\hspace{-5mm}
= \int d2\,
\bar{V}(1,2)
 \langle T_{\tau}\bar{\psi}(2)\psi(2)\psi(1)\bar{\phi}(1')\rangle ,
\label{G-Dyson}
\end{eqnarray}
\begin{eqnarray}
&&\hspace{-2mm}
-\biggl(\frac{\partial}{\partial \tau_{1}}-K_{1}\biggr)\bar{G}(1,1')+\delta(1,1')
\nonumber \\
&&\hspace{-5mm}
=\int d2\,
\bar{V}(1,2)
\langle T_{\tau}\bar{\psi}(1)\bar{\psi}(2)\psi(2)\phi(1')\rangle .
\label{G^d-Dyson}
\end{eqnarray}
\end{subequations}
We then set $1'=1_{+}$ and $1'=1_{-}$ in Eqs.\ (\ref{G-Dyson}) and (\ref{G^d-Dyson}), respectively. 
We also multiply Eqs.\ (\ref{Psi-eq-motion}) and (\ref{Psi^d-eq-motion}) 
with $\eta_{\rm ext}=0$ by $\bar{\Psi}(1)$ and $\Psi(1)$ from the left, respectively.
Let us add the four equations, 
perform an integration over $1$, and multiply the result by $T/4$ with
Eq.\ (\ref{H_int}), Eq.\ (\ref{psi=Psi+psi-t}), and 
$\bar{V}(1,1')\equiv \delta(\tau_{1}-\tau_{1}')V({\bf r}_{1}-{\bf r}_{1}')$ 
in mind.
We thereby obtain
\begin{eqnarray}
&&\hspace{-5mm}
\langle H_{\rm int}\rangle
=\frac{T}{4}
\int d1 
\biggl\{ \bar{\Psi}(1)\biggl(-\frac{\partial}{\partial \tau_{1}}-K_{1}\biggr)\Psi(1)
\nonumber \\
&&\hspace{8.8mm}
+\Psi(1)\biggl(\frac{\partial}{\partial \tau_{1}}-K_{1}\biggr)\bar{\Psi}(1)
\nonumber \\
&&\hspace{8.8mm}
+\biggl[-\biggl(-\frac{\partial}{\partial \tau_{1}}-K_{1}\biggr)G(1,1')
+\delta(1,1')\biggr]_{1'=1_{+}}
\nonumber \\
&&\hspace{8.8mm}
+\biggl[-\biggl(\frac{\partial}{\partial \tau_{1}}\!-\! K_{1}\biggr)\bar{G}(1,1')
+\delta(1,1')\biggr]_{1'=1_{-}}\!\biggr\} .
\nonumber \\
\label{<H_int>A}
\end{eqnarray}
We subsequently express the right-hand side of Eq.\ (\ref{<H_int>A}) in terms of the self-energies
by using Eq.\ (\ref{Dyson}), Eq.\  (\ref{GP}), $\bar{G}(1,2)=G(2,1)$, and $\bar{\Sigma}(1,2)=\Sigma(2,1)$.
Noting that the subscript in $G(1_{-},2)$ is effective only for 
$\tau_{2}=\tau_{1}$ where $G(1_{-},2)=G(1,2_{+})$, we arrive at Eq.\ (\ref{<H_int>}).

\section{Expression of $\Phi^{(3)}$}

The basic diagrams for $\Phi^{(3)}$ are given in Fig.\ \ref{fig:Phi^(3)}.
Inserting arrows into them in all possible ways, we obtain  $81$ distinct diagrams. The corresponding $\Phi^{(3)}$ may be expressed compactly as
\begin{widetext}
\begin{eqnarray}
&&\hspace{-5mm}
\Phi^{(3)}
=
\frac{T}{24}{\rm Tr}\bigl[ 4
\bar{G}G\Gamma^{(0)}\bar{G}G\Gamma^{(0)}\bar{G}G\Gamma^{(0)}
+GG\Gamma^{(0)}GG\Gamma^{(0)}GG\Gamma^{(0)}
+c^{(3)}_{6c}\bar{F}F\Gamma^{(0)}\bar{G}G\Gamma^{(0)}\bar{G}G\Gamma^{(0)}
\nonumber \\
&&\hspace{6mm}
+c^{(3)}_{6d}\bar{F}G\Gamma^{(0)}FG\Gamma^{(0)}\bar{G}G\Gamma^{(0)}
+c^{(3)}_{6e}FG\Gamma^{(0)}\bar{F}G\Gamma^{(0)}GG\Gamma^{(0)}
+c^{(3)}_{6f}\bar{F}F\Gamma^{(0)}\bar{F}F\Gamma^{(0)}\bar{G}G\Gamma^{(0)}
\nonumber \\
&&\hspace{6mm}
+c^{(3)}_{6g}\bar{F}F\Gamma^{(0)}\bar{F}G\Gamma^{(0)}FG\Gamma^{(0)}
+c^{(3)}_{6h}FF\Gamma^{(0)}\bar{F}\bar{F}\Gamma^{(0)}GG\Gamma^{(0)}
+c^{(3)}_{6i}(FF\Gamma^{(0)}\bar{F}\bar{G}\Gamma^{(0)}\bar{F}G\Gamma^{(0)}+{\rm c.c.})
\nonumber \\
&&\hspace{6mm}
+c^{(3)}_{6j}\bar{F}F\Gamma^{(0)}\bar{F}F\Gamma^{(0)}\bar{F}F\Gamma^{(0)}
+c^{(3)}_{5a}\bar{g}G\Gamma^{(0)}\bar{G}G\Gamma^{(0)}\bar{G}G\Gamma^{(0)}
+c^{(3)}_{5b}gG\Gamma^{(0)}GG\Gamma^{(0)}GG\Gamma^{(0)}
\nonumber \\
&&\hspace{6mm}
+c^{(3)}_{5c}(\bar{f}F+{\rm c.c.})\Gamma^{(0)}\bar{G}G\Gamma^{(0)}\bar{G}G\Gamma^{(0)}
+c^{(3)}_{5d}(\bar{f}G\Gamma^{(0)}FG\Gamma^{(0)}\bar{G}G\Gamma^{(0)}+{\rm c.c.})
+c^{(3)}_{5e}(FG\Gamma^{(0)}\bar{f}G\Gamma^{(0)}GG\Gamma^{(0)}+{\rm c.c.})
\nonumber \\
&&\hspace{6mm}
+c^{(3)}_{5f}\bar{F}F\Gamma^{(0)}\bar{G}G\Gamma^{(0)}(\bar{g}G+{\rm c.c.})\Gamma^{(0)}
+c^{(3)}_{5g}\bar{F}G\Gamma^{(0)}FG\Gamma^{(0)}\bar{g}G\Gamma^{(0)}
+c^{(3)}_{5h}\bar{F}G\Gamma^{(0)}FG\Gamma^{(0)}\bar{G}g\Gamma^{(0)}
\nonumber \\
&&\hspace{6mm}
+c^{(3)}_{5i}(\bar{F}g\Gamma^{(0)}FG\Gamma^{(0)}\bar{G}G\Gamma^{(0)}+{\rm c.c.})
+c^{(3)}_{5j}FG\Gamma^{(0)}\bar{F}G\Gamma^{(0)}gG\Gamma^{(0)}
+c^{(3)}_{5k}(FG\Gamma^{(0)}\bar{F}g\Gamma^{(0)}GG\Gamma^{(0)}+{\rm c.c.})
\nonumber \\
&&\hspace{6mm}
+c^{(3)}_{5\ell}(\bar{f}F+{\rm c.c.})\Gamma^{(0)}\bar{F}F\Gamma^{(0)}\bar{G}G\Gamma^{(0)}
+c^{(3)}_{5m}(\bar{f}F+{\rm c.c.})\Gamma^{(0)}\bar{F}G\Gamma^{(0)}FG\Gamma^{(0)}
+c^{(3)}_{5n}(\bar{F}F\Gamma^{(0)}\bar{f}G\Gamma^{(0)}FG\Gamma^{(0)}+{\rm c.c.})
\nonumber \\
&&\hspace{6mm}
+c^{(3)}_{5o}(FF\Gamma^{(0)}\bar{f}\bar{F}\Gamma^{(0)}GG\Gamma^{(0)}+{\rm c.c.})
+c^{(3)}_{5p}(fF\Gamma^{(0)}\bar{F}\bar{G}\Gamma^{(0)}\bar{F}G\Gamma^{(0)}+{\rm c.c.})
+c^{(3)}_{5q}(FF\Gamma^{(0)}\bar{f}\bar{G}\Gamma^{(0)}\bar{F}G\Gamma^{(0)}+{\rm c.c.})
\nonumber \\
&&\hspace{6mm}
+c^{(3)}_{5r}\bar{F}F\Gamma^{(0)}\bar{F}F\Gamma^{(0)}\bar{g}G\Gamma^{(0)}
+c^{(3)}_{5s}(\bar{F}F\Gamma^{(0)}\bar{F}g\Gamma^{(0)}FG\Gamma^{(0)}+{\rm c.c.})
+c^{(3)}_{5t}FF\Gamma^{(0)}\bar{F}\bar{F}\Gamma^{(0)}gG\Gamma^{(0)}
\nonumber \\
&&\hspace{6mm}
+c^{(3)}_{5u}(FF\Gamma^{(0)}\bar{F}\bar{G}\Gamma^{(0)}\bar{F}g\Gamma^{(0)}+{\rm c.c.})
+c^{(3)}_{5v}(\bar{f}F+{\rm c.c.})\Gamma^{(0)}\bar{F}F\Gamma^{(0)}\bar{F}F\Gamma^{(0)}
+c^{(3)}_{4a}\bar{g}G\Gamma^{(0)}\bar{g}G\Gamma^{(0)}\bar{G}G\Gamma^{(0)}
\nonumber \\
&&\hspace{6mm}
+c^{(3)}_{4b}gG\Gamma^{(0)}gG\Gamma^{(0)}GG\Gamma^{(0)}
+c^{(3)}_{4c}(\bar{g}G\Gamma^{(0)}\bar{G}g\Gamma^{(0)}\bar{G}G\Gamma^{(0)}+{\rm c.c.})
+c^{(3)}_{4d}(\bar{f}F\Gamma^{(0)}\bar{G}g\Gamma^{(0)}\bar{G}G\Gamma^{(0)}+{\rm c.c.})
\nonumber\\
&&\hspace{6mm}
+c^{(3)}_{4e}(\bar{f}F\Gamma^{(0)}\bar{g}G\Gamma^{(0)}\bar{G}G\Gamma^{(0)}+{\rm c.c.})
+c^{(3)}_{4f}(\bar{f}G\Gamma^{(0)}FG\Gamma^{(0)}\bar{g}G\Gamma^{(0)}+{\rm c.c.})
+c^{(3)}_{4g}(\bar{f}G\Gamma^{(0)}FG\Gamma^{(0)}\bar{G}g\Gamma^{(0)}+{\rm c.c.})
\nonumber \\
&&\hspace{6mm}
+c^{(3)}_{4h}(Fg\Gamma^{(0)}\bar{f}G\Gamma^{(0)}GG\Gamma^{(0)}+{\rm c.c.})
+c^{(3)}_{4i}\bar{F}g\Gamma^{(0)}Fg\Gamma^{(0)}\bar{G}G\Gamma^{(0)}
+c^{(3)}_{4j}\bar{F}F\Gamma^{(0)}\bar{g}G\Gamma^{(0)}\bar{g}G\Gamma^{(0)}
\nonumber \\
&&\hspace{6mm}
+c^{(3)}_{4k}(\bar{F}F\Gamma^{(0)}\bar{G}g\Gamma^{(0)}\bar{g}G\Gamma^{(0)}+{\rm c.c.})
+c^{(3)}_{4\ell}(\bar{F}G\Gamma^{(0)}Fg\Gamma^{(0)}\bar{g}G\Gamma^{(0)}+{\rm c.c.})
+c^{(3)}_{4m}(\bar{F}g\Gamma^{(0)}FG\Gamma^{(0)}\bar{G}g\Gamma^{(0)}+{\rm c.c.})
\nonumber \\
&&\hspace{6mm}
+c^{(3)}_{4n}(FG\Gamma^{(0)}\bar{F}g\Gamma^{(0)}gG\Gamma^{(0)}+{\rm c.c.})
+c^{(3)}_{4o}Fg\Gamma^{(0)}\bar{F}g\Gamma^{(0)}GG\Gamma^{(0)}
+c^{(3)}_{4p}(FF\Gamma^{(0)}\bar{f}\bar{G}\Gamma^{(0)}\bar{f}G\Gamma^{(0)}+{\rm c.c.})
\nonumber \\
&&\hspace{6mm}
+c^{(3)}_{4q}(\bar{f}F\Gamma^{(0)}\bar{f}G\Gamma^{(0)}FG\Gamma^{(0)}+{\rm c.c.})
+c^{(3)}_{4r}(\bar{f}F\Gamma^{(0)}\bar{f}F\Gamma^{(0)}\bar{G}G\Gamma^{(0)}+{\rm c.c.})
+c^{(3)}_{4s}(\bar{f}F\Gamma^{(0)}\bar{F}F\Gamma^{(0)}\bar{g}G\Gamma^{(0)}+{\rm c.c.})
\nonumber \\
&&\hspace{6mm}
+c^{(3)}_{4t}(\bar{f}F\Gamma^{(0)}\bar{F}F\Gamma^{(0)}\bar{G}g\Gamma^{(0)}+{\rm c.c.})
+c^{(3)}_{4u}(\bar{f}F\Gamma^{(0)}\bar{F}g\Gamma^{(0)}FG\Gamma^{(0)}+{\rm c.c.})
+c^{(3)}_{4v}(\bar{f}F\Gamma^{(0)}\bar{F}G\Gamma^{(0)}Fg\Gamma^{(0)}+{\rm c.c.})
\nonumber \\
&&\hspace{6mm}
+c^{(3)}_{4w}(FF\Gamma^{(0)}\bar{f}\bar{F}\Gamma^{(0)}gG\Gamma^{(0)}+{\rm c.c.})
+c^{(3)}_{4x}\bar{F}F\Gamma^{(0)}\bar{F}g\Gamma^{(0)}Fg\Gamma^{(0)}
+c^{(3)}_{4y}(\bar{f}F\Gamma^{(0)}\bar{f}F\Gamma^{(0)}\bar{F}F\Gamma^{(0)}+{\rm c.c.})
\nonumber \\
&&\hspace{6mm}
+c^{(3)}_{4z}(FF\Gamma^{(0)}\bar{F}\bar{g}\Gamma^{(0)}\bar{F}g\Gamma^{(0)}+{\rm c.c.})
+c^{(3)}_{4\alpha}\bar{g}g\Gamma^{(0)}\bar{G}G\Gamma^{(0)}\bar{G}G\Gamma^{(0)}
+c^{(3)}_{4\beta}gg\Gamma^{(0)}GG\Gamma^{(0)}GG\Gamma^{(0)}
\nonumber \\
&&\hspace{6mm}
+c^{(3)}_{4\gamma}(\bar{f}g\Gamma^{(0)}FG\Gamma^{(0)}\bar{G}G\Gamma^{(0)}+{\rm c.c.})
+c^{(3)}_{4\delta}(FG\Gamma^{(0)}\bar{f}g\Gamma^{(0)}GG\Gamma^{(0)}+{\rm c.c.})
+c^{(3)}_{4\varepsilon}\bar{F}F\Gamma^{(0)}\bar{G}G\Gamma^{(0)}\bar{g}g\Gamma^{(0)}
\nonumber \\
&&\hspace{6mm}
+c^{(3)}_{4\kappa}\bar{F}G\Gamma^{(0)}FG\Gamma^{(0)}\bar{g}g\Gamma^{(0)}
+c^{(3)}_{4\lambda}FG\Gamma^{(0)}\bar{F}G\Gamma^{(0)}gg\Gamma^{(0)}
+c^{(3)}_{4\mu}(ff\Gamma^{(0)}\bar{F}\bar{G}\Gamma^{(0)}\bar{F}G\Gamma^{(0)}+{\rm c.c.})
\nonumber \\
&&\hspace{6mm}
+c^{(3)}_{4\nu}(FF\Gamma^{(0)}\bar{f}\bar{f}\Gamma^{(0)}GG\Gamma^{(0)}+{\rm c.c.})
+c^{(3)}_{4\pi}(\bar{F}F\Gamma^{(0)}\bar{f}g\Gamma^{(0)}FG\Gamma^{(0)}+{\rm c.c.})
+c^{(3)}_{4\rho}(FF\Gamma^{(0)}\bar{f}\bar{g}\Gamma^{(0)}\bar{F}G\Gamma^{(0)}+{\rm c.c.})
\nonumber \\
&&\hspace{6mm}
+c^{(3)}_{4\sigma}\bar{F}F\Gamma^{(0)}\bar{F}F\Gamma^{(0)}\bar{g}g\Gamma^{(0)}
+c^{(3)}_{4\tau}FF\Gamma^{(0)}\bar{F}\bar{F}\Gamma^{(0)}gg\Gamma^{(0)}
+c^{(3)}_{3a}\bar{g}G\Gamma^{(0)}\bar{g}G\Gamma^{(0)}\bar{g}G\Gamma^{(0)}
\nonumber \\
&&\hspace{6mm}
+c^{(3)}_{3b}\bar{g}G\Gamma^{(0)}\bar{g}G\Gamma^{(0)}\bar{G}g\Gamma^{(0)}
+c^{(3)}_{3c}(\bar{f}F\Gamma^{(0)}\bar{G}g\Gamma^{(0)}\bar{g}G\Gamma^{(0)}+{\rm c.c.})
+c^{(3)}_{3d}(\bar{f}F\Gamma^{(0)}\bar{g}G\Gamma^{(0)}\bar{G}g\Gamma^{(0)}+{\rm c.c.})
\nonumber
\end{eqnarray}
\begin{eqnarray}
&&\hspace{6mm}
+c^{(3)}_{3e}(\bar{f}F\Gamma^{(0)}\bar{g}G\Gamma^{(0)}\bar{g}G\Gamma^{(0)}+{\rm c.c.})
+c^{(3)}_{3f}\bar{F}g\Gamma^{(0)}Fg\Gamma^{(0)}\bar{G}g\Gamma^{(0)}
+c^{(3)}_{3g}\bar{F}g\Gamma^{(0)}Fg\Gamma^{(0)}\bar{g}G\Gamma^{(0)}
\nonumber \\
&&\hspace{6mm}
+c^{(3)}_{3h}(\bar{f}F\Gamma^{(0)}\bar{f}G\Gamma^{(0)}Fg\Gamma^{(0)}+{\rm c.c.})
+c^{(3)}_{3i}(\bar{f}F\Gamma^{(0)}\bar{F}g\Gamma^{(0)}Fg\Gamma^{(0)}+{\rm c.c.})
+c^{(3)}_{3j}(\bar{f}F\Gamma^{(0)}\bar{f}F\Gamma^{(0)}\bar{f}F\Gamma^{(0)}+{\rm c.c.}) \bigr] .
\nonumber \\
\label{Phi^(3)}
\end{eqnarray}
Here $\bar{G}G$ etc.\ connect adjacent two vertices with appropriately chosen arguments as
\begin{eqnarray}
&&\hspace{-6mm}
{\rm Tr}
\bar{G}G\Gamma^{(0)}\bar{G}G\Gamma^{(0)}\bar{G}G\Gamma^{(0)}
=\int d1\cdots \int d6'\,\bar{G}(1',6)G(1,6')
\Gamma^{(0)}(66',55')\bar{G}(5',4)G(5,4')\Gamma^{(0)}(44',33')
\nonumber \\
&&\hspace{36.6mm}\times\bar{G}(3',2)G(3,2')\Gamma^{(0)}(22',11'),
\end{eqnarray}
\begin{eqnarray}
&&\hspace{-6mm}
{\rm Tr}
FF\Gamma^{(0)}\bar{F}\bar{G}\Gamma^{(0)}\bar{F}G\Gamma^{(0)}
=\int d1\cdots \int d6'\,F(1,5)F(2,6)
\Gamma^{(0)}(66',55')\bar{F}(6',4')\bar{G}(5',4)\Gamma^{(0)}(44',33')
\nonumber \\
&&\hspace{36.6mm}\times\bar{F}(3',1')G(3,2')\Gamma^{(0)}(22',11'),
\end{eqnarray}
and $g$, $\bar{g}$, $f$, and $\bar{f}$ are composed of $\Psi$ and $\bar{\Psi}$ as 
\begin{equation}
g(1,2)=\Psi(1)\bar{\Psi}(2),\hspace{5mm}\bar{g}(1,2)=\bar{\Psi}(1)\Psi(2),
\hspace{5mm}
f(1,2)=\Psi(1)\Psi(2),\hspace{5mm}\bar{f}(1,2)=\bar{\Psi}(1)\bar{\Psi}(2) .
\end{equation}
The unknown coefficients in Eq.\ (\ref{Phi^(3)}) have been determined by Eq.\ (\ref{relation1})
using the diagrams of Figs.\ \ref{fig:Phi^(3)} and  \ref{fig:Phi^(3)-A}.
The final results are summarized as follows:
\begin{subequations}
\label{c^(3)}
\begin{eqnarray}
&&\hspace{-10mm}
c^{(3)}_{6c}=-15,\hspace{5mm}c^{(3)}_{6d}=c^{(3)}_{6g}=-c^{(3)}_{6e}=6,\hspace{5mm}
c^{(3)}_{6f}=12,\hspace{5mm}
c^{(3)}_{6h}=-c^{(3)}_{6i}=3,\hspace{5mm}
c^{(3)}_{6j}=-5,
\label{c^(3)_6-summary}
\end{eqnarray}
\begin{eqnarray}
&&\hspace{-10mm}
c^{(3)}_{5b}=c^{(3)}_{5d}=c^{(3)}_{5g}=c^{(3)}_{5h}=c^{(3)}_{5i}=c^{(3)}_{5m}=c^{(3)}_{5n}=c^{(3)}_{5o}=c^{(3)}_{5s}=c^{(3)}_{5t}=-6,\hspace{5mm}
c^{(3)}_{5e}=c^{(3)}_{5k}=c^{(3)}_{5p}=c^{(3)}_{5q}=c^{(3)}_{5u}=6,\hspace{5mm}
\nonumber \\
&&\hspace{-10mm}
c^{(3)}_{5a}=c^{(3)}_{5l}=c^{(3)}_{5r}=-24,\hspace{5mm}
c^{(3)}_{5c}=c^{(3)}_{5v}=15,\hspace{5mm}
c^{(3)}_{5f}=30,\hspace{5mm}
c^{(3)}_{5j}=12,
\label{c^(3)_5-summary}
\end{eqnarray}
\begin{eqnarray}
&&\hspace{-4mm}
c^{(3)}_{4a}=c^{(3)}_{4i}=c^{(3)}_{4s}=-c^{(3)}_{4e}=30,\hspace{5mm}
c^{(3)}_{4c}=c^{(3)}_{4r}=c^{(3)}_{4\kappa}=12,\hspace{5mm}
c^{(3)}_{4d}=c^{(3)}_{4j}=c^{(3)}_{4x}=-c^{(3)}_{4t}=-24,\hspace{5mm}
c^{(3)}_{4k}=c^{(3)}_{4y}=-15,
\nonumber \\
&&\hspace{-4mm}
c^{(3)}_{4b}=c^{(3)}_{4g}=c^{(3)}_{4l}=c^{(3)}_{4o}=c^{(3)}_{4q}=c^{(3)}_{4v}=c^{(3)}_{4w}=c^{(3)}_{4\gamma}=c^{(3)}_{4\pi}=6,\hspace{5mm}
c^{(3)}_{4p}=c^{(3)}_{4z}=c^{(3)}_{4\alpha}=c^{(3)}_{4\sigma}=c^{(3)}_{4\mu}=-3,
\nonumber \\
&&\hspace{-4mm}
c^{(3)}_{4f}=c^{(3)}_{4h}=c^{(3)}_{4m}=c^{(3)}_{4n}=c^{(3)}_{4u}=c^{(3)}_{4\delta}
=c^{(3)}_{4\varepsilon}=c^{(3)}_{4\lambda}=c^{(3)}_{4\rho}=-6,\hspace{5mm}
c^{(3)}_{4\beta}=c^{(3)}_{4\nu}=c^{(3)}_{4\tau}=3,\hspace{5mm}
\label{c^(3)_4-summary}
\end{eqnarray}
\begin{equation}
c^{(3)}_{3a}=-10 ,\hspace{5mm}
c^{(3)}_{3b}=c^{(3)}_{3f}=c^{(3)}_{3g}=c^{(3)}_{3h}=-c^{(3)}_{3e}=-30,\hspace{5mm}
c^{(3)}_{3c}=c^{(3)}_{3d}=c^{(3)}_{3i}=15,\hspace{5mm}
c^{(3)}_{3j}=5.
\label{c^(3)-3-summary}
\end{equation}
\end{subequations}
\end{widetext}
It has been confirmed that Eq.\ (\ref{Phi^(3)}) with Eq.\ (\ref{c^(3)}) also satisfies Eq.\ (\ref{relation2}).

\end{document}